\documentclass[twocolumn, a4paper]{article}

\usepackage{cite}
\usepackage{amsmath,amssymb,amsfonts}
\usepackage{amsthm}
\usepackage{gensymb}
\usepackage{algorithm}
\usepackage{algpseudocode}
\usepackage[colorlinks]{hyperref}
\usepackage[noabbrev,capitalize,nameinlink]{cleveref}
\usepackage{graphicx}
\usepackage{textcomp}
\usepackage{multicol}
\usepackage{empheq}
\usepackage{authblk}
\usepackage[left=1cm, right=1cm, top=2cm]{geometry} 
\providecommand{\keywords}[1]
{
  \small	
  \textbf{\textit{Keywords---}} #1
}

\usepackage{subfigure}

\newcommand{\W}{\mathbb{W}}
\newcommand{\N}{\mathbb{N}}
\newcommand{\LHH}{L^{HH}}
\newcommand{\sHAP}{\textit{serving} HAP}
\newcommand{\sFSO}{\emph{serving} FSO}

\newcommand{\M}{\mathbb{M}}
\newcommand{\V}{\mathbb{V}}
\newcommand{\Hh}{\mathbb{H}}
\newcommand{\MHAP}{\mathbb{M}^{HAP}}
\newcommand{\nsFSOi}{n^{\textrm{sF}}_i} 
\newcommand{\niFSOi}{n^{\textrm{iF}}_i}

\newcommand{\NFSO}{\N_{\FSO}}

\newcommand{\CdayH}{\zeta^{day}_\textrm{H}}
\newcommand{\CdayF}{\zeta^{day}_\textrm{F}}
\newcommand{\Ccharge}{\zeta^{mtn}}
\newcommand{\Pngang}{\rho^{inter}_{\textrm{F}}}
\newcommand{\PngangHi}{P^{inter}_{H_i}}

\newcommand{\PdownHi}{P_{H_i}^{down}}
\newcommand{\Pheat}{\rho^{\textrm{HCM}}_{\textrm{F}}}
\newcommand{\Ppat}{\rho^{\textrm{PAT}}}
\newcommand{\PtxFSO}{\rho^{\FSO}_{\textrm{tx}}}
\newcommand{\mFSO}{\mu_{\textrm{F}}}
\newcommand{\mPlatform}{\mu_{\textrm{H}}}
\newcommand{\Phap}{P^{avion}_{H_i}}
\newcommand{\Pflyunit}{\rho^{avion}}
\newcommand{\Prx}{\rho_{\textrm{rx}}}

\newcommand{\Rrx}{\mathbb{R}_{\textrm{rx}}}
\newcommand{\Psolar}{\mathbb{E}^{solar}}
\newcommand{\Econsum}{\mathbb{E}^{consum}}
\newcommand{\Dm}{\mathbb{D}^{m}}
\newcommand{\HAP}{\textrm{HAP}}
\newcommand{\FSO}{\textrm{FSO}}
\newcommand{\Linter}{\textrm{L}^{inter}}

\newcommand{\at}{\frac{\alpha}{2}}

\newcommand{\bt}{\frac{\beta}{2}}
\newcommand{\pim}{\frac{\pi}{m}}
\newcommand{\axt}{\frac{\xi + \alpha}{2}}
\newcommand{\CostO}{\widehat{Cost}}
\newcommand{\Cost}{Cost}
\newcommand{\KO}{\hat{K}}

\newcommand{\Ralpha}{R_{\alpha}}

\newcommand{\Rext}{R_{\textrm{ext}}}

\newcommand{\msupp}{$m$FSO}
\newcommand{\msuppconf}{\msupp~configuration}

\newtheorem{theorem}{Theorem}
\newtheorem{lemma}[theorem]{Lemma}

\title{Optimal multiple FSO transceiver configuration for using on High-altitude platforms}
\author[1]{Dieu Linh Truong \thanks{linhtd@soict.hust.edu.vn}}
\author[2]{The Ngoc Dang \thanks{ngocdt@ptit.edu.vn}}
\affil[1]{School of Information and Communication Technology, Hanoi University of Science and Technology, Vietnam}
\affil[2]{Department of Wireless Communications, Posts and Telecommunication Institute of Technology, Vietnam}

\begin{document}
\maketitle
\begin{abstract}
Free-space optical (FSO) communication requires light of sight (LoS) between the transmitter and the receiver. For long-distance communication, many research projects have been conducted towards using a network composed of high-altitude platforms (HAPs) flying at an elevation of 20 km to carry intermediate FSO transceivers that forward data between ground stations. The clear environment at high elevations prevents terrestrial obstacles from cutting the LoS between the transceivers. An FSO transceiver on a HAP can communicate with ground stations within a small area owing to its limited beam size. We suggest using multiple FSO transceivers on a HAP to extend its ground coverage. However, the use of too many FSO transceivers may quickly exhaust the onboard energy of the HAP. As a result, HAP must be lowered to recharge frequently.  

In this study, we first propose a configuration of multiple FSO transceivers to widen the ground coverage of a HAP. We then propose a set of closed-form expressions to calculate the extended coverage.  Finally, to implement a HAP network using multiple FSO transceivers, we seek the optimal configuration of multiple FSO transceivers that minimizes the total cost of the HAP network, including amortization, energy, and maintenance costs. The simulation results show that the proposed multiple FSO transceiver configuration clearly increases  the ground coverage of a HAP and significantly reduces the cost of the HAP network.
\end{abstract}


\section{Introduction}
Free-space optical (FSO) communication uses light propagation in free space to transmit data. In recent years, this technology has emerged as a promising choice for short-distance high-speed communication between endpoints with a clear light of sight (LoS). Commercial FSO transmitters available in the market at prices of thousands of dollars can operate at $1.25-10$ Gbps over $1-2$ kilometers, for example, the SONABeam series of fSona \cite{fSona}. 

To reach a long distance, a multi-hop FSO system can be used, where data are transmitted through intermediate FSO transceivers \cite{FirstFSOnet1999}, \cite{ZangFSOnet2002}. To avoid obstacles that cut the LoS between terrestrial FSO transceivers, researchers from academia and industry have proposed placing intermediate FSO transceivers of the multi-hop FSO system on high-altitude platforms (HAPs). High-altitude platforms are flying objects that operate at altitudes of 17–24 km in the stratosphere. Several HAP models have been proposed and piloted previously. Some projects continue until recently, such as the Loon Project of Google \cite{LoonProject}, the UAV project of Facebook \cite{FB2017}, and the Stratobus project of Thales Alenia Space \cite{Stratobus}.

\begin{figure}[tbh]
    \centering
    \includegraphics[width=0.5\textwidth]{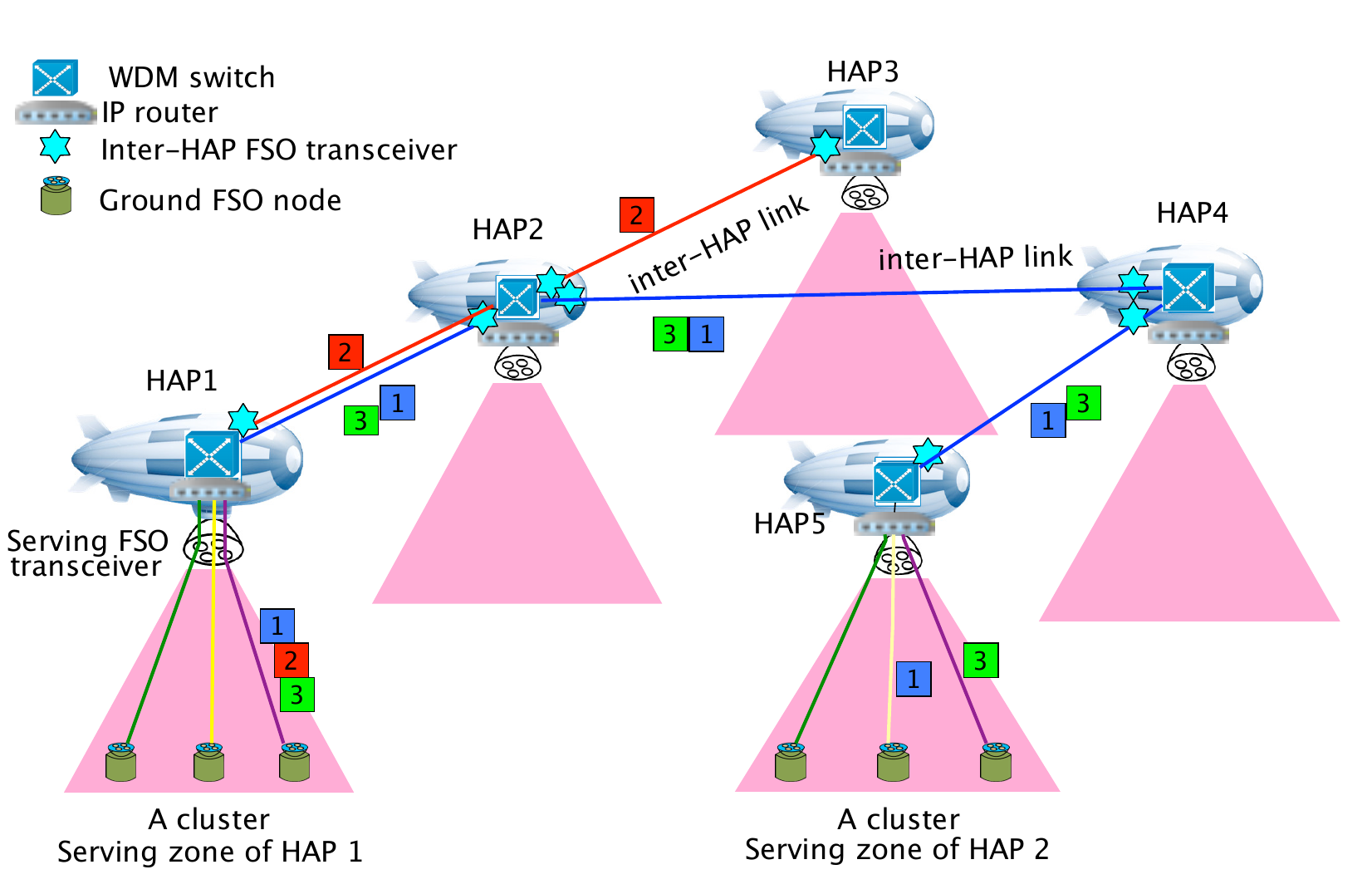}
    \caption{Multi-hop FSO communication system using HAP.}
    \label{fig:comm-model}
\end{figure}

A multi-hop FSO system using a HAP network is described in \cite{osnpaper} and illustrated in Figure \ref{fig:comm-model}. According to this model, FSO transceivers on the ground (so-called ground FSO nodes) are regrouped into clusters to become  the serving zones of HAPs. A HAP has an FSO transceiver looking down to exchange data with the ground FSO nodes of the cluster under it. This FSO transceiver is called \sFSO~transceiver. A HAP also carries several FSO transceivers pointing towards other HAPs for inter-HAP communication. These FSO transceivers are known as \textit{inter-HAP} FSO transceivers. 

Although the ITU recommends a HAP footprint width of approximately 500 km in radius, experimental projects show much smaller coverage areas \cite{framework2021}. Nevertheless, a network of multiple HAPs can cover a country entirely. For example, a constellation of 16 HAPs with multiple radio frequency antennas  was considered to cover Japan  \cite{Miura2001}.

An end-to-end data-switching scheme for a multi-hop FSO system using HAP was proposed in \cite{osnpaper}. Since the communication between a HAP and the ground is point-to-multipoint, the \sFSO~ transceiver on the HAP controls multiple accesses from ground FSO nodes under it using the WDM technique. Each ground node is assigned a separate wavelength for up and down communication. An IP router on the HAP aggregates IP packets heading toward a common cluster within a single flow.  The flow will be carried by one or more continuous lightpaths between the source and destination HAPs. The number of lightpaths is determined according to the size of the flow and the transport capacity of a wavelength. A WDM switch is installed on each HAP to route these lightpaths over the HAP network on a wavelength-switched basis. In Figure \ref{fig:comm-model}, the blue path HAP1-HAP2-HAP4-HAP5 and the red path HAP1-HAP2-HAP3 are two flows. 

In terrestrial FSO communications, the light beams are usually set to be very narrow for low transmission energies. However, for HAP and ground communication,  the \sFSO~transceiver of the HAP must  project a sufficiently wide laser beam for covering distributed ground FSO nodes.

A single \sFSO~transceiver has a relatively small footprint owing to the low capacity of the current laser source, and the limited sensibility and aperture sizes of ground receivers. The calculation in Section \ref{sec:single-conf} shows that with a laser source of 1 Watt, required received power at receivers of -41.1dBm, and receiver aperture radius of 2 m, a single \sFSO~transceiver at an elevation of 20 km can cover a ground area of 6.691 km radius only (see Table \ref{tab:max-alpha}). 
\begin{figure}
\centering
\includegraphics[width=0.4\textwidth]{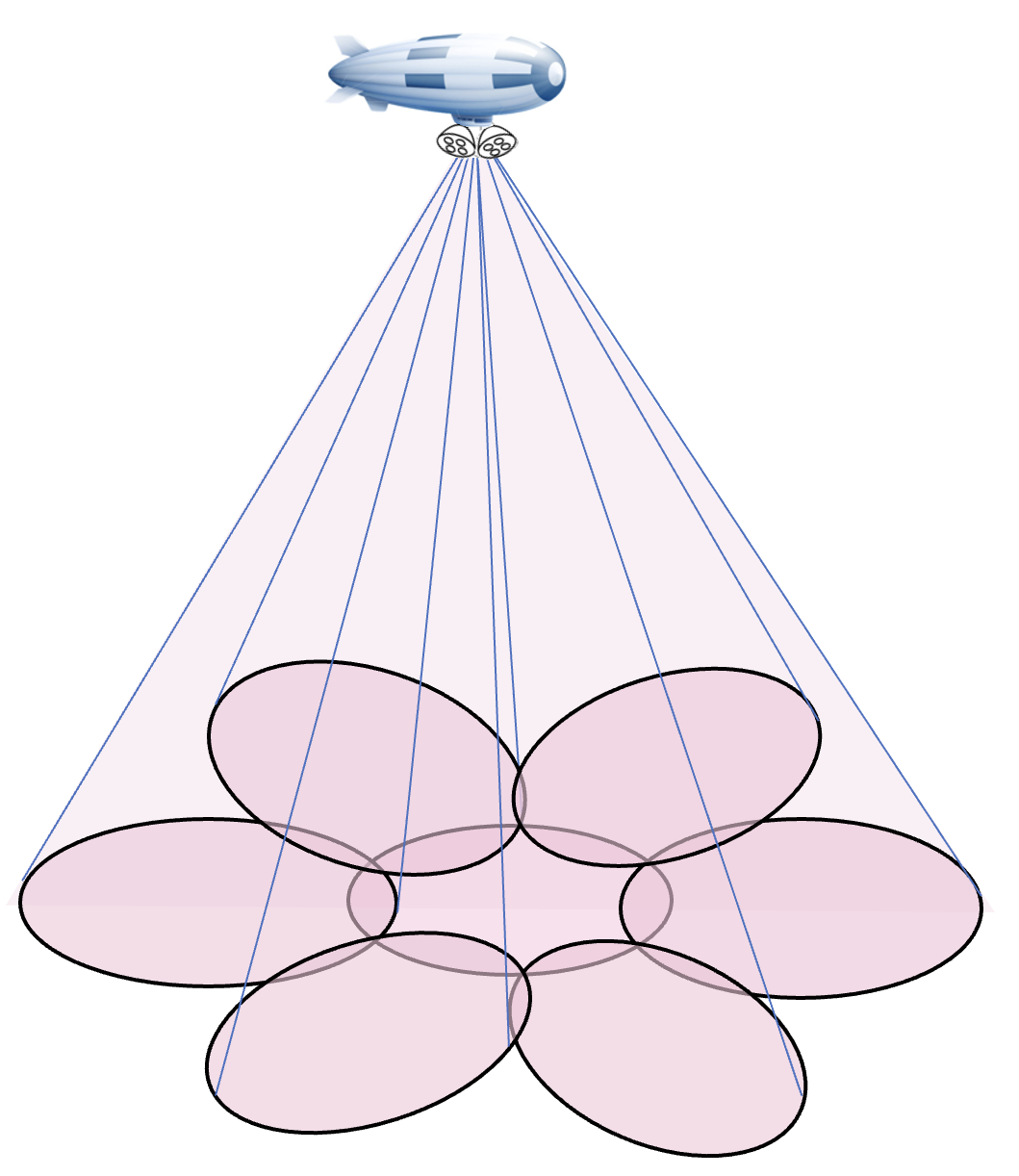}
\caption{A HAP with multiple \sFSO~transceivers and its footprint.}
\label{fig:HAP-multiFSO}
\end{figure}


To extend the coverage of a HAP, we propose using multiple \sFSO~transceivers arranged in a bundle, as shown in Figure \ref{fig:HAP-multiFSO}. Each \sFSO~transceiver points in a slightly different direction to cover a particular ground area that overlaps other areas to create a continuous coverage region. Given a ground region to be served, using HAPs with multiple \sFSO~transceivers reduces the number of required HAPs compared to using HAPs with a single \sFSO~transceiver. However, the expenditure for \sFSO~transceivers increases. Therefore, the number of \sFSO~transceivers to be used on a HAP should be carefully considered. 

Regarding the communication between ground nodes and a HAP, the multiple \sFSO~transceiver model still uses the WDM technique, where each ground node is assigned a unique wavelength within its cluster to communicate with its HAP. The number of ground nodes to be served by a HAP is restricted by the number of wavelengths offered by the WDM technique. 

In this study, we focus on identifying the optimal configuration of multiple \sFSO~transceivers to achieve a minimal-cost HAP network for serving a set of ground FSO nodes. The optimal configuration should define the number of \sFSO~transceivers to be set up on a HAP and the beam width for each transceiver. The cost of the HAP network includes the investment, energy, and maintenance costs.

Compared with the previous study in reference \cite{osnpaper}, the current research differs in two aspects. First, the current research proposes the use of multiple \sFSO~transceivers on each HAP instead of a single \sFSO~transceiver, as in  \cite{osnpaper}. Second, the current research identifies the optimal beam widths for \sFSO~transceivers, whereas in \cite{osnpaper}, the beam widths are predefined.

The current study also differs from that in \cite{Optimal-beam-size-2019}, where beam size was optimized for an inter-HAP link, which is a point-to-point link.

The remainder of this paper is organized as follows. First, we analyze the single and multiple \sFSO~transceivers configurations in Section \ref{sec:optimal-radius} to determine their ground coverage sizes and constraints on transmitter beams. In Section \ref{sec:scope}, we state  the problem of designing a minimal-cost HAP-based FSO network, which is the target of the optimization of multiple \sFSO~transceiver configuration. Then, in Section \ref{sec:energy}, we define a HAP energy consumption formula and show that solar energy is necessary for keeping the HAP working in space for a long period. We also present a constraint that a HAP must respect to relying uniquely on solar energy. In Section \ref{sec:opt1}, we present the algorithms for identifying the optimal multiple \sFSO~transceiver configuration and its footprint radius. Section \ref{sec:algo} presents the process designing the minimal cost HAP-based FSO network using the optimal multiple \sFSO~transceiver configuration.  Section \ref{sec:experiment} presents the simulation results. Finally, Section \ref{sec:conclusions} concludes the paper.

\section{Serving FSO transceiver configurations}
\label{sec:optimal-radius}

\subsection{Single serving FSO transceiver configuration}
\label{sec:single-conf}
In this section, the allowable beam width and ground coverage of a single \sFSO~transceiver are determined. The beam size is restricted to ensure that the received power at a ground point within the beam footprint is detectable by receivers.  
 \begin{figure}
   \includegraphics[width=0.5 \textwidth]{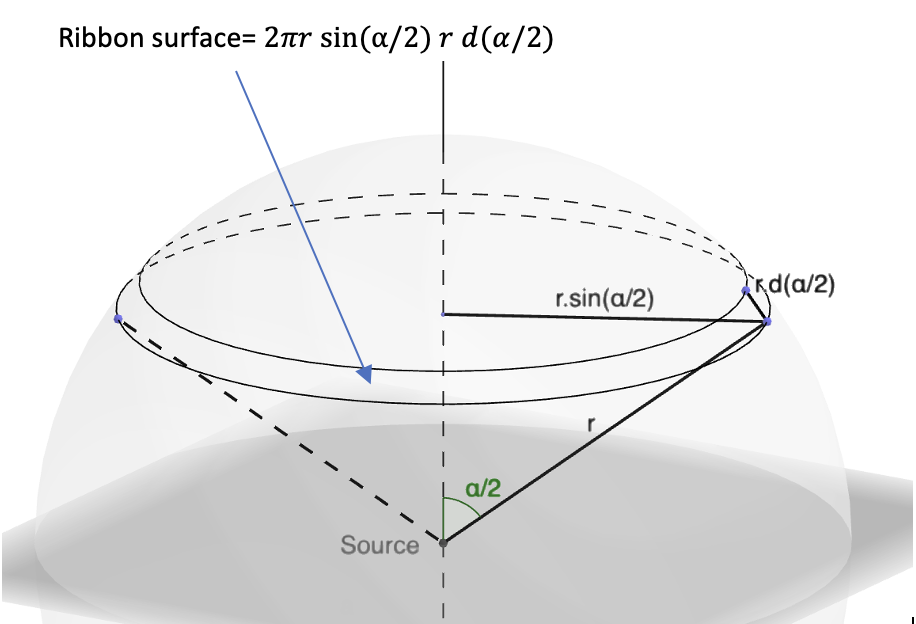}
    \caption{Surface of the part of sphere blocked by solid angle $\alpha$ is calculated as the sum of the surface of all ribbons around the sphere when the solid angle varies from $\alpha$ to 0.}
    \label{fig:radiation-density}
\end{figure}

Assume that the transmitter source radiates within a solid angle $\alpha$ and that the radiation density is uniform in all directions within the solid angle at a distance $r$ from the source. The radiation density at distance $r$ is inversely proportional to the surface of the part of the sphere radius $r$ blocked by the solid angle $\alpha$. To calculate this surface, we divide the sphere into thin ribbons corresponding to open angles of $d(\alpha/2)$. The width of a ribbon is $r d(\alpha/2)$, as shown in Figure \ref{fig:radiation-density}. The radius of the ribbon at zenith angle $\alpha/2$ is $r\sin(\alpha/2)$. Thus, the ribbon surface is $2 \pi r \sin(\alpha/2) r d(\alpha/2)$. The surface of the part of the sphere blocked by the solid angle $\alpha$ is the sum of the surfaces of all ribbons when zenith angle varies from $\alpha$ to 0, as follows:
$$\int_{\alpha}^0 2 \pi r \sin{(\frac{\alpha}{2})} r d (\frac{\alpha}{2}) = 2\pi r^2 (1-\cos{(\frac{\alpha}{2})})$$

Let $U_r$ be the radiation density at distance $r$ and $P_{tx}$ be the transmitted power at the source. We deduce:

\begin{equation}
U_r = \frac{P_{tx}}{2 \pi r^2 (1-\cos{(\alpha/2)})}
\label{eq:U-r}
\end{equation}

\begin{figure}
   \includegraphics[width=0.5 \textwidth]{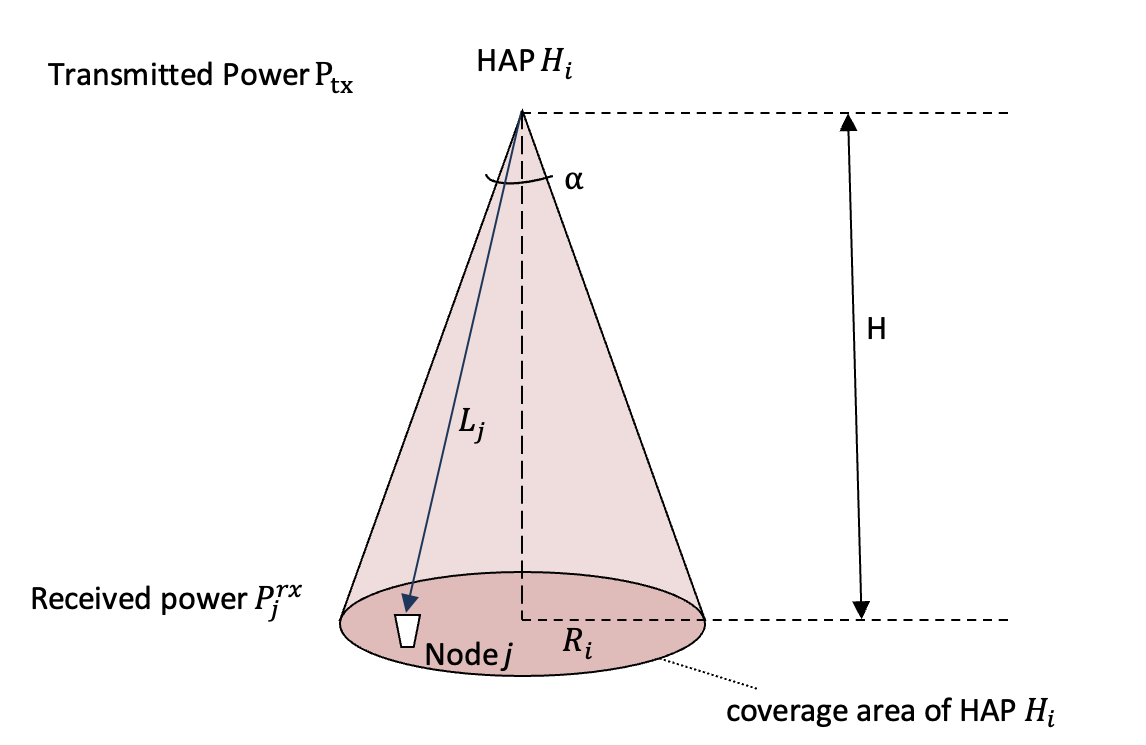}
    \caption{Received power on border nodes of a coverage area is the smallest amongst all nodes in the area.}
    \label{fig:Prx-alpha}
\end{figure}

Let $P^{rx}_j$ be the received power at ground FSO node $j$. The received power is proportional to the radiation density and the received aperture of the ground node. It is:
\begin{eqnarray}
	P^{rx}_j &=& e^{-\sigma L_j} U_{L_j} A_{R} 
	\label{eq:Pdown-Prx}
\end{eqnarray}

where
\begin{itemize}
	\item $L_j$ is the distance between ground FSO node $j$ and its \sHAP~$H_i$ (see Figure \ref{fig:Prx-alpha}),
	\item $\sigma$ is the attenuation coefficient of the links between the HAP and ground,
	\item $U_{L_j}$ is radiation density at distance $L_j$ from the source,
	\item $A_R$  is the aperture area of the receiver. Let $\Rrx$ be the receiver aperture radius, then, $A_R=\pi \Rrx^2$.
\end{itemize}
In \eqref{eq:Pdown-Prx}, the first term represents the attenuation of laser power through the atmosphere, which is described by the exponential Beer–Lambert Law \cite{PrxFormula}. 

By substituting $U_{L_j}$ from \eqref{eq:U-r} into \eqref{eq:Pdown-Prx}, we obtain the received power at node $j$ as follows:
\begin{equation}
    P^{rx}_j= e^{-\sigma L_j} \times P_{tx} \times \frac{\Rrx^2}{2 L_j^2} \times \frac{1}{1-\cos{(\alpha/2)}} 
    \label{eq:power-J}
\end{equation}
The power received at node $j$ must not be less than the required level of the receiver, denoted by $\Prx$. It is obvious that point $j$ at the border of the ground coverage area receives the least power because it is the furthest from the source (see Figure \ref{fig:Prx-alpha}). Hence, all points in the coverage areas of HAP $H_i$ receive sufficient power if and only if the border points receive at least the required power; that is,

\begin{equation}
\boxed{
P^{rx}_j=e^{-\sigma \Hh/ \cos{(\frac{\alpha}{2})}}  \frac{P_{tx} \Rrx^2 \cos^2{(\frac{\alpha}{2})}}{2 \Hh^2  (1-\cos{(\frac{\alpha}{2})})} \geq \Prx
}
\label{eq:alpha-max}
\end{equation}
where $L_j$ is substituted by $\Hh/\cos(\frac{\alpha}{2})$ for border node $j$.

Solving inequation \eqref{eq:alpha-max} yields the beam width of the single \sFSO~transceiver configuration. Corresponding to beam width $\alpha$, the ground coverage radius of the configuration is:
\begin{equation}
R_i = \Hh \tan(\frac{\alpha}{2})
\end{equation}

\begin{figure}[tbh]
    \centering
    \includegraphics[width=0.5\textwidth]{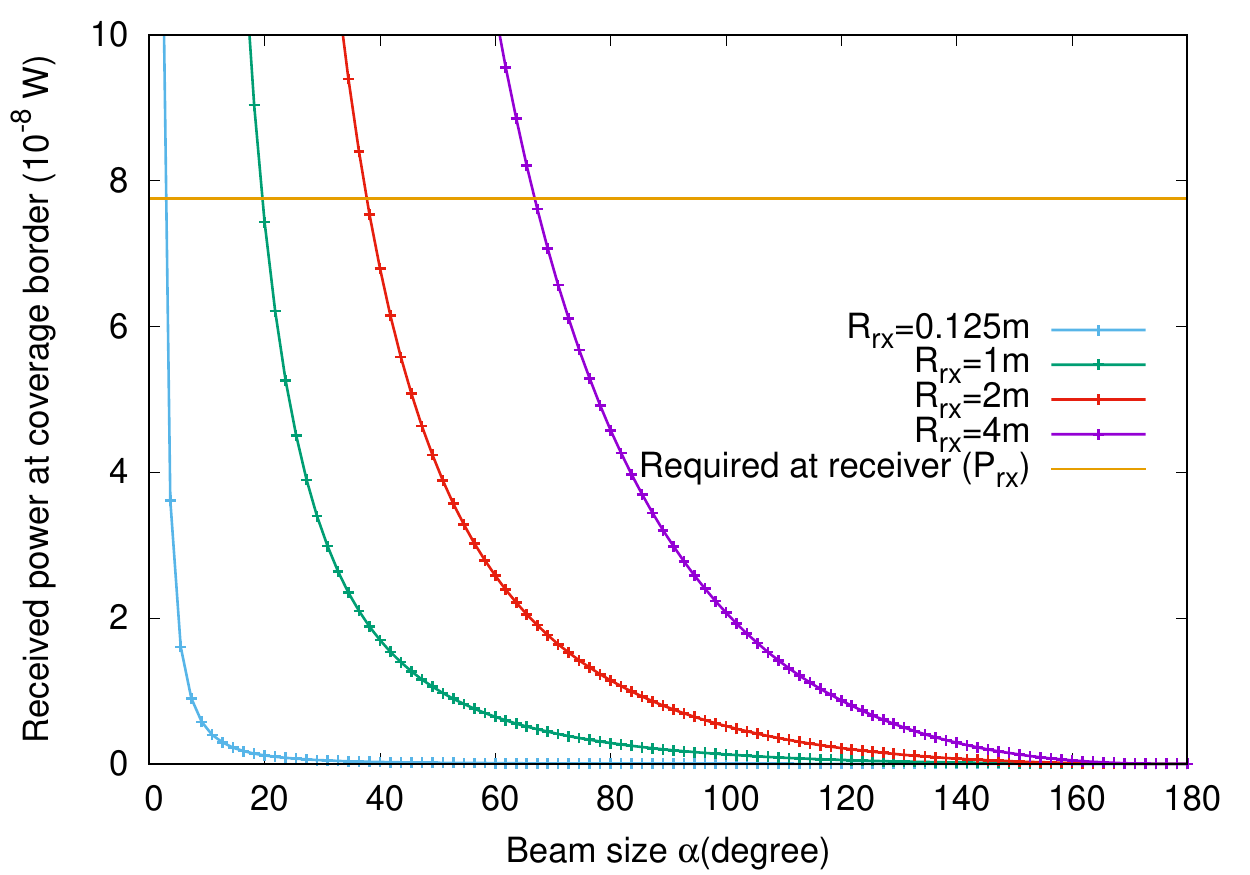}
    \caption{Received power at the coverage border of the single \sFSO~transceiver configuration with different receiver apertures. }
    \label{fig:alpha-max}
\end{figure}

\begin{lemma}
Function $P^{rx}_j$ decreases with $\alpha \in [0 ..\pi]$. 
\label{lemma1}
\end{lemma}
Proof of Lemma \ref{lemma1} is given in Appendix \ref{proof-lemma1}.

Figure \ref{fig:alpha-max} shows the received power at the border of the coverage area with different receiver aperture radius $\Rrx$. 
This figure confirms that $P^{rx}_j$ decrease with an increase in $\alpha$. 

Let $\alpha_{max}$ be the value for $\alpha$ that makes  $P^{rx}_j(\alpha_{max})=\Prx$; then according to Lemma \ref{lemma1}, 
$$P^{rx}_j(\alpha) \geq P^{rx}_j(\alpha_{max})=\Prx, \forall \alpha \in [0 .. \alpha_{max}]$$
thus all $\alpha \in [0 .. \alpha_{max}]$ satisfy constraint \eqref{eq:alpha-max}. 

Calculations using the parameters given in Table \ref{tab:param} show that when $\Rrx=2$ m, $\alpha_{max}=37\degree$ and the coverage radius is 6.691~km. When $\Rrx=4$ m, $\alpha_{max}=67\degree$ and the coverage radius is 13.237~km.
 
\subsection{Multiple serving FSO transceiver configuration}

\begin{figure}[tbh]
\centering
   \includegraphics[width=0.4 \textwidth]{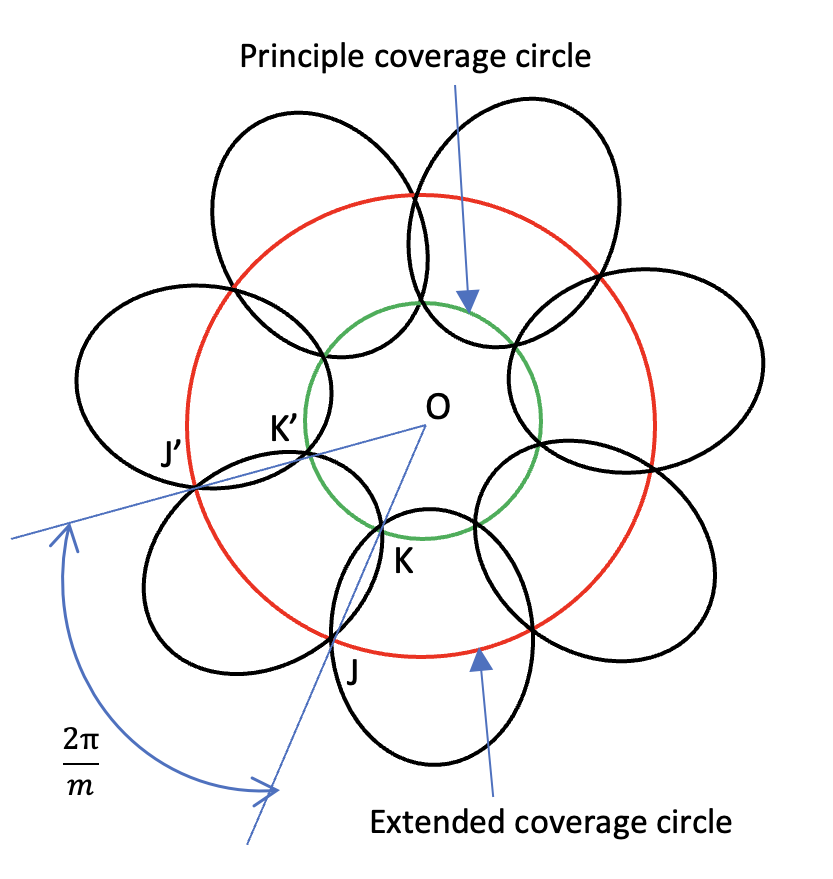}
    \caption{footprint of multiple FSO transceiver (\msupp) configuration.}
    \label{fig:mFSOconf}
\end{figure}

The ground coverage of a HAP can be widened by combining several \sFSO~transceivers. Different combinations are possible. In this research, we study a straightforward configuration in which a principal \sFSO~transceiver is in the center projecting light perpendicular to the ground, and several identical supplementary \sFSO~transceivers are set evenly around the principal one (Figure \ref{fig:mFSOconf}). Each supplementary transceiver projects slanted beams to extend the coverage in one direction. This arrangement is referred to as \msuppconf. Usually, the transmitters in a bundle are considered to project signals in parallel. However, because of the large principal beam, the supplementary \sFSO~transceiver projection directions are far from being perpendicular to the ground, and their footprints are ellipses instead of circles. 

To create a continuous coverage region, the footprint of the principal \sFSO~transceiver and those of the supplementary \sFSO~transceivers should overlap. Therefore, there should be a sufficiently large number of supplementary \sFSO~transceivers to cover entirely the contour of the principal footprint. The extended coverage area is defined as the largest circle covered by these footprints (Figure \ref{fig:mFSOconf}). The principal transceiver is responsible for the region defined by its footprint. A supplementary \sFSO~transceiver is responsible for the part limited by its footprint, principal coverage circle, and extended coverage circle. 

Let $\alpha$ be always the beam width of the principal \sFSO~transceiver. To ensure that ground nodes under principal coverage receive sufficient power, $\alpha$ should still respect constraint \eqref{eq:alpha-max}, as in the single \sFSO~transceiver configuration. 

Let the beam width of a supplementary \sFSO~transceiver be $\beta$. In the responsible area of the supplementary transceiver, the points on the extended coverage circle are the farthest from the supplementary transceiver; thus, they receive the least power. If these points receive at least $\Prx$, all other points receive sufficient power.

It is easy to note that the footprints of the neighboring supplementary \sFSO~transceivers join each other on the extended coverage circle. Let $J$ be such a joint point, the power $J$ receives from the supplementary FSO transceiver is defined similar to \eqref{eq:power-J} but with beam width $\beta$, which is
\begin{equation}
    P^{rx}_J= e^{-\sigma \times L_J} \times P_{tx} \times \frac{\Rrx^2}{4 L_J^2} \times \frac{2}{1-\cos{(\beta/2)}} 
    \label{eq:Prx-beta}
\end{equation}

Thus, $\beta$ is constrained by the condition $P^{rx}_J \geq \Prx$, which gives:
\begin{equation}
\boxed{
    e^{-\sigma \times L_J}  \frac{P_{tx}\Rrx^2}{2L_J^2(1-\cos{(\beta/2))}} \geq \Prx
    }
    \label{eq:beta-max}
\end{equation}

Let us denote the extended coverage radius by $\Rext$ then

\begin{equation}
\boxed{
L_J= \sqrt{\Hh^2 + \Rext^2}
}
\label{eq:Ljbeta-1}
\end{equation}

Appendix \ref{sec:calculation} presents detailed calculations of $L_J$ and $\Rext$. The calculations yielded the following results
\begin{equation}
\boxed{
\Rext=\Hh \frac{2\tan(\axt) -\tan(\at) (1- \tan^2(\axt))}{1 - \tan^2(\axt) + 2 \tan(\axt).\tan(\at)}
}
\label{eq:r-extended-1}
\end{equation}
where 
\begin{empheq}[box=\fbox]{align}
    \tan(\axt) =&\frac{\tan(\gamma) + \tan(\theta)}{1-\tan(\gamma).\tan(\theta)}. \cos(\frac{\pi}{m})\label{eq:axt}\\
    \tan(\gamma)=& \tan(\frac{\alpha}{2}).\cos(\frac{\pi}{m}) \label{eq:tan-gamma}\\
    \tan(\theta)=&\frac{\sqrt{\sin^2(\frac{\beta}{2})  -\sin^2(\frac{\alpha}{2}).\sin^2(\frac{\pi}{m})}}{\cos(\frac{\beta}{2})} \label{eq:tan-phi}
\end{empheq}
and $m$ is the number of supplementary FSO transceivers set around the principal one.

We can remark that $\Rext$ and thus $L_J$ depend on $\alpha$, $\beta$ and $m$. Hereafter, $\Rext$ is sometimes denoted by $\Rext(\alpha, m, \beta)$ and $L_J$ by $L_J(\alpha, m, \beta)$ to express these dependencies.

\section{Problem of designing minimal cost HAP network}
\label{sec:scope}

\begin{table*}[tbh]
\begin{tabular}{l | l | l | l}
\hline
\parbox{1.2cm}{Param. notations} & Descriptions & Values 		&References \\
\hline\multicolumn{3}{c}{Cost related parameters} \\
\hline
$\CdayH$& Daily amortization cost of a HAP.  & 100		&			\\
$\CdayF$& Daily amortization cost of an FSO transceiver on HAP. &	 10		&			\\
$\Ccharge$	& Cost of one-time maintenance of a HAP  including lowing it down, &	1000		&			\\
&maintenance, charging and reinstall it in the stratosphere. &&\\
$\Dm$   &Maintenance cycle. & 365 days & \cite{Stratobus} \\
\hline
\multicolumn{3}{c}{Energy parameters} \\
\hline
$\Psolar$		& Minimum daily harvested solar energy by a HAP. & 42 - 290 kWh	& \cite{solar-energyHAP2020}	\\
$\Pflyunit$ 	& Power consumed by the avionic part of a HAP to carry an unit of mass. & 2 W/kg & \\
$\Pheat$ & Power for heating, cooling, and management for each FSO on HAP. & 20 W & \cite{LoonProject}\\
$\Ppat$ & Power consumed by a PAT system. & 15W & \cite{Optical-HAP-2010}\\
$\Pngang$ 	& Power consumed by inter-HAP FSO transceivers for laser source (0.1 W), & & \\
& heating/cooling/management (20 W) and PAT (15 W). & 35.1 W & \cite{LoonProject}\\
\hline
\multicolumn{3}{c}{Inter-HAP FSO link parameters} \\
\hline
$C_n^2$ & Atmosphere structure parameter. & $5.0 \times 10^{-18} m^{-2/3}$ &\\
- &Attenuation coefficient. & $3.5 \times 10^{-6} m^{-1}$ & \cite{LoonProject} \\
- & Coupling loss.& 45 dBm & \\

- & Transmitted power of an inter-HAP FSO transceiver. & 0.1 W & \cite{LoonProject} \\
- & Receiver aperture diameter of an inter-HAP FSO transceiver. & 0.037 m & \cite{LoonProject} \\
- & Beam width of an inter-HAP FSO transmitter. & 280 $\mu$rad & \cite{LoonProject} \\
\hline
\multicolumn{3}{c}{HAP-ground link parameters and variables} \\
\hline
$\sigma$ & Attenuation coefficient. &  $3.5 \times 10^{-6} m^{-1}$ &\\
$\PtxFSO$ & Transmitted power of the laser source of a \sFSO~transceiver. & 1 Watt & \\
$\Rrx$ & Receiver aperture radius of a ground FSO transceiver. & 0.05 m & SONABeam \cite{fSona}\\
$\Prx$		& Required received power at a ground FSO transceiver. &  $7.76.10^{-8}$ W& -41.1 dBm in \cite{LoonProject} \\
\hline
\multicolumn{3}{c}{Other parameters} \\
\hline
$\Hh$ & Elevation of HAPs. & 20 km & \\
$\LHH$		&  Maximum length of an inter-HAP link so that its BER is under $\delta$. & 88 km   \\
$\delta$    & BER threshold for inter-HAP links and lightpaths between HAPs.			\\
$\W$	&	The number of wavelengths in WDM technique. &40; 80		&			\\
$\mPlatform$	& Platform mass excluding FSO transceivers. &28.5 kg; 500 kg 		& \cite{LoonProject}\\
$\mFSO$		& FSO transceiver mass.   &6.3 kg 		& \cite{LoonProject} \\
\hline 
\end{tabular}
\caption{Parameters. Greek characters are used for denoting constant parameters.}
\label{tab:param} 
\end{table*}
\color{black}

There are several costs in a HAP network, such as investment, energy, and maintenance costs. Based on the expected life duration and maintenance cycle of a HAP,  these costs can be distributed by day as 1) daily amortization cost representing investment cost, 2) average daily maintenance cost, and 3) daily energy cost. Consequently, the problem of minimizing network cost becomes minimizing the daily network cost, which comprises these three components.
 
Following variables are introduced for formulating mathematically the daily network cost: 
\begin{itemize}
	\item $K$: Number of HAPs in the network. The HAPs are indexed by $i \in 1..K$.
	\item $\niFSOi$: Number of FSO transceivers used on $\HAP_i$ for inter-HAP communications.
	\item $\nsFSOi$: Number of \sFSO~transceiver of $\HAP_i$. 
\end{itemize}

Let $\CdayH$ and $\CdayF$ be constants that express the daily amortization costs of a HAP and an FSO transceiver, respectively. These costs are defined as the ratio of the prices of the HAP or FSO transceiver to their expected lifetime duration. Then, the overall \emph{daily amortization cost} of the HAP network is:
\begin{equation}
	K  \CdayH + (\sum_{i=1}^K \nsFSOi + \sum_{i=1}^K \niFSOi) \CdayF
	\label{eq:cost-amort}
\end{equation}

To evaluate the daily maintenance and energy costs, we need to consider the HAP design. HAPs are classified into two categories based on the underlying physical principle that provides the lifting force for the HAPs: aerodynamic (the HAP is heavier than air) and aerostatic (the HAP is lighter than air). While aerostatic platforms use buoyancy to float in the air, aerodynamic platforms use dynamic forces created by movement through the air \cite{framework2021}. In general, both aerostatic and aerodynamic systems require a ``flying energy'' to keep the HAP  relatively stable for maintaining FSO communication between HAPs and that between HAPs and FSO ground nodes. An aerodynamic system requires a large propulsion power to move. Aerostatic systems typically consume less energy than aerodynamic systems do. To be able to operate for a long duration in space, HAPs are mainly unmanned.

HAPs are equipped with different energy resources such as onsite production (e.g., solar energy harvested by solar panels) or rechargeable energy (e.g., batteries or fuel cells brought from the ground). Solar energy-based HAPs can operate continuously in space until they are lowered for maintenance purpose. Rechargeable energy-based HAPs are lowered once the reserved energy is depleted. In brief, the continuous in-space working duration of a HAP is limited by its available energy, which is relatively fixed by the HAP design, its energy consumption level, which varies depending on the payload weight and communication of the HAP, and its maintenance cycle.

We define the \emph{maintenance cost} of a HAP as the expense of lowering the HAP to perform technical maintenance, energy recharge on the ground, and then reinstall it in space. 

Let $d_i$ be the number of days on which $\HAP_i$ can operate continuously in space. Let $\Ccharge$ be constant expressing the cost of one time lowering a HAP, maintaining it, recharging it, and then reinstalling it in space. The \textit{daily maintenance cost} of the HAP network is
\begin{equation}
	 \sum_{i=1}^K \frac{\Ccharge}{d_i}
	\label{eq:cost-mtn}
\end{equation}

Regarding the \emph{daily energy cost}, we consider solar energy to be free, whereas the solar panel cost is counted in the cost of the HAP. The cost of rechargeable energy is part the maintenance cost. As a result, the energy cost does not explicitly represent the total cost. Nonetheless, the energy consumption level of a HAP affects its in-space working duration  $d_i$; therefore, we analyze this in Section \ref{sec:energy}.

Combining \eqref{eq:cost-amort} and \eqref{eq:cost-mtn}, we obtain the following overall daily cost of the HAP network:
\begin{equation}
	\Cost=K \CdayH + (\sum_{i=1}^K \nsFSOi + \sum_{i=1}^K \niFSOi )\CdayF + \sum_{i=1}^K \frac{\Ccharge}{d_i}
	\label{cost}
\end{equation}

The problem of minimizing daily cost of the HAP network is stated as follows.
\begin{itemize}
\item Given input parameters including
\begin{itemize}
	\item $\NFSO$: Set of ground FSO nodes and their coordinates. The number of nodes in the set is denoted as $|\N_{\FSO}|$,
	\item $\M$: Data traffic to be carried between ground FSO nodes. This is the list of bandwidth demands between the ground nodes.
\end{itemize}
\item Outputs to seek are
\begin{itemize}
	\item A HAP network with HAP locations and inter-HAP links,
	\item Beam width to set to each \sFSO~transceiver.
\end{itemize}
\item Optimization objective is
\begin{itemize}
    \item Minimizing the daily cost expressed in \eqref{cost} of the HAP network. 
\end{itemize}
\end{itemize}

The following two remarks drive us to conduct further analyses in subsequent sections. First, if a HAP has self-sufficient solar energy, its in-space working duration $d_i$ is not limited by its energy consumption but depends uniquely on the maintenance cycle of the HAP, which is usually constant. In Section \ref{sec:energy}, we show the daily energy consumption of a HAP and the constraint that a HAP needs to respect to rely solely on solar energy.

Second, the cost of the HAP network increases with an increase in the number of FSO transceivers and HAPs. The number of HAPs can be reduced by increasing ground coverage. To increase ground coverage, more \sFSO~transceivers can be used on each HAP, but this introduces greater energy consumption and extra amortization cost.  Section \ref{sec:opt1} focuses on identifying the optimal configuration for \sFSO~transceivers on a HAP to achieve a minimal HAP network cost.
\section{Daily energy consumption of a HAP with payload}
\label{sec:energy}

Several parameters affect the power consumption of a HAP. The descriptions and notations of these parameters are listed in section Energy parameters of Table  \ref{tab:param}. Most parameters were set based on industrial experimental projects such as the Loon project \cite{LoonProject}, Stratobus project \cite{Stratobus}, and other studies listed in the reference column. Section \ref{sec:param} presents the choice of parameter values in detail.

Let us consider the power consumption of a single HAP $H_i$ that has $m$ \sFSO~transceiver and $\niFSOi$ inter-HAP FSO transceivers. The power consumption includes:
\begin{itemize}
	\item $\Phap$: Power draw of avionic part for maintaining $H_i$ with payload in space. 
	\item $\PdownHi$: Power draw of all \sFSO~transceivers on HAP $H_i$. This power includes the heating/cooling/management power, laser transmitted power of all \sFSO~transceivers on the HAP, and the power consumed by the Pointing Acquisition and Tracking (PAT) system of the HAP. 
	\item $\PngangHi$: Power draw of all inter-HAP FSO transceivers on HAP $H_i$ for inter-HAP communication. The power includes the heating/cooling/management, and PAT power for each inter-HAP FSO transceiver. Inter-HAP FSO transceivers are oriented towards different remote HAPs; therefore, each transceiver must have a PAT system.
\end{itemize}

The total daily energy consumption (by 24 hours) of $H_i$ is
\begin{equation}
\Econsum= (\Phap + \PdownHi + \PngangHi) \times 24
\label{eq:Etotal}
\end{equation}
To breakdown further $\Phap$, $\PdownHi$, and $\PngangHi$, we introduce following parameters:

\begin{itemize}
	\item $\Pflyunit$: Power consumed by the avionic part of the HAP to carry a unit of mass.
	\item  $\PtxFSO$: Transmitted power of each \sFSO~transceiver. Because the current power of laser source is limited to 1~W, which is very small in comparison with the power consumed by other factors on the HAP, we consider that $\PtxFSO=1$ W, regardless of the beam width of the \sFSO~transceiver. 
	\item $\Pheat$: Power draw for heating, cooling, and management. It is also considered constant for each \sFSO~transceiver and is set to $\Pheat=20$ W, according to reference \cite{LoonProject}.
	\item $\Ppat$ : Power draw for Pointing, Acquisition and Tracking activity; it is another constant and is set to $\Ppat=15$ W \cite{Optical-HAP-2010}. A HAP system uses a single PAT for its set of \sFSO~transceivers. 
	\item $\Pngang$: Power draw of a single inter-HAP FSO transceiver including communication, heating, cooling, management, and PAT. According to \cite{LoonProject}, 0.1~W laser power is sufficient for an inter-HAP communication of 100~km distance. In this study, we limited the inter-HAP link length to less than 100 km and considered the laser power constantly 0.1~W regardless of the distance. Therefore, $\Pngang=\Pheat +\Ppat +0.1$.
	\item $\mPlatform$:  Mass of the HAP.
	\item $\mFSO$: Mass of an FSO on the HAP.
\end{itemize}

Assuming that $\Phap$ is linearly proportional to the weight of the HAP by $\Pflyunit$, 
\begin{equation}
	\Phap =  [\mPlatform + (\nsFSOi+ \niFSOi) \mFSO]  \Pflyunit
	\label{eq:pHAP}
\end{equation}

$\PdownHi$ is the sum of the power consumed by \sFSO~transceivers and PAT activity of the HAP; thus,
\begin{equation}
	\PdownHi =  \nsFSOi  (\PtxFSO+ \Pheat) + \Ppat
	\label{eq:down}
\end{equation}

$\PngangHi$ is the sum of the power consumed by inter-HAP FSO transceivers; thus,
\begin{equation}
\PngangHi=\Pngang . \niFSOi
	\label{eq:pngang}
\end{equation}

Substituting  \eqref{eq:pHAP}, \eqref{eq:down}, and \eqref{eq:pngang} into \eqref{eq:Etotal}, we obtain the daily power consumption of a HAP as
\begin{equation}
\boxed{
\begin{aligned}
\Econsum &= \{ [\mPlatform + (\nsFSOi+ \niFSOi) \mFSO]  \Pflyunit \\
 &+ \nsFSOi  (\PtxFSO+ \Pheat) + \Ppat \\
 &+ \Pngang  \niFSOi \} \times 24
 \end{aligned}}
 \label{eq:Etotal-detail}
\end{equation}

\subsection{Necessity of solar energy and utilization constraint }
\label{sec:need-solar}
Current HAPs mainly use energy from solar panels mounted on HAP wings and/or energy from batteries or hydrogen fuel cells (HFC) onboard. Solar energy can be harvested and charged into batteries during the day for nighttime use. Harvested solar energy varies with year time and location. According to the experiments in \cite{solar-energyHAP2020}, in York, UK, the harvested solar power is 42--80 kWh/day, and in Enugu, Nigeria, it is 290--545 kWh/day, depending on the size of the solar panel. 

\begin{figure}
   \includegraphics[width=0.45\textwidth]{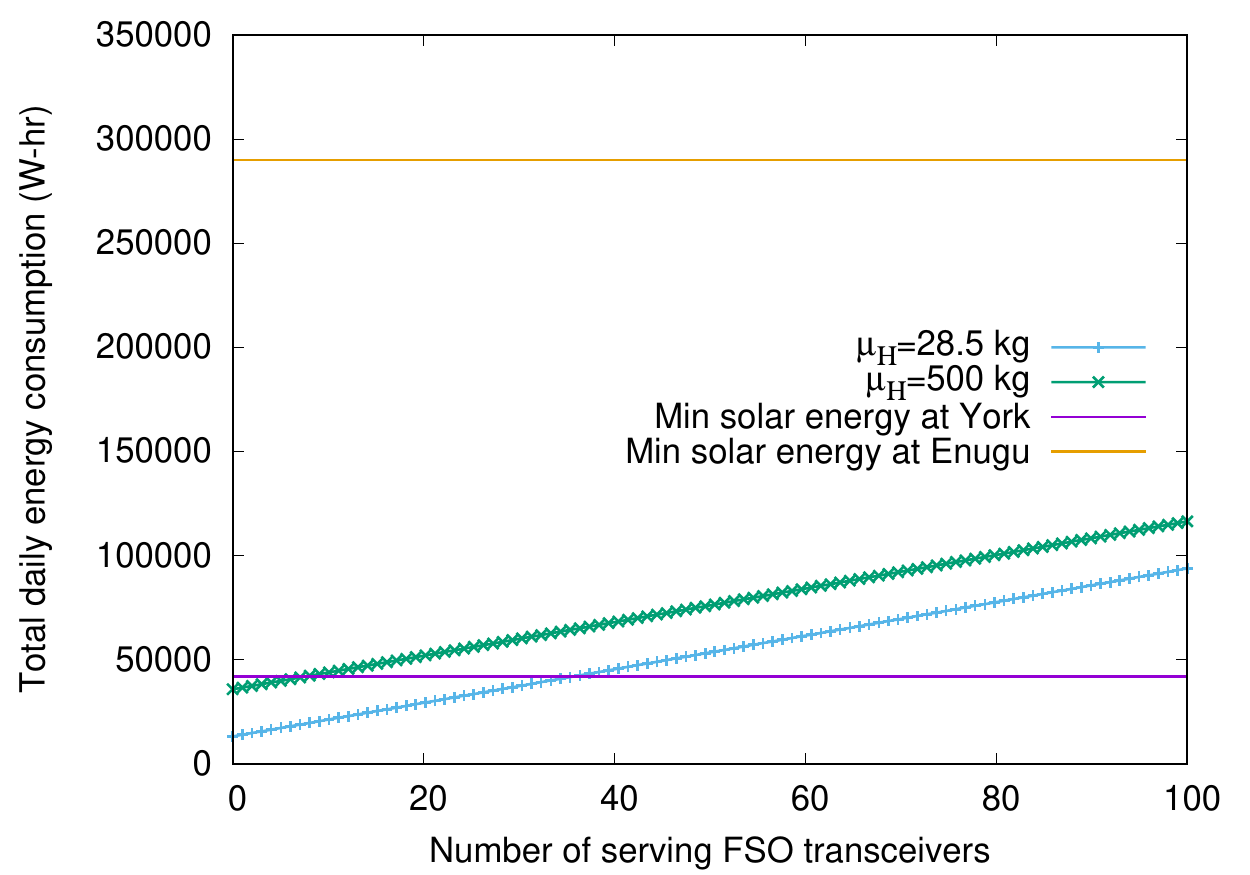}
    \caption{Energy consumption by a HAP with different number of serving FSO transceivers in comparison with the minimum harvested solar energy at York and Enugu. $\Pflyunit=2/kg$~W and $\Ppat = 15$~W.}
    \label{fig:energy-beam}
\end{figure}

Figure \ref{fig:energy-beam} depicts the total daily energy consumption of a HAP, calculated from \eqref{eq:Etotal-detail}, versus the number of \sFSO~transceivers. Parameters were $\Pflyunit=2$~W/kg, $\Ppat = 15$~W, HAP weights $\mPlatform=28.5$ kg or 500 kg. The HAP carried 10 inter-HAP FSO transceivers. The referenced daily solar energy levels were the minimum daily solar energy levels in York and Enugu. From a certain number of \sFSO~transceivers, a HAP consumes more energy than the harvested solar energy in York, and an HFC would be necessary. Owing to the limited payload capacity of a HAP, its HFC capacity is also very limited. According to \cite{framework2021}, the current state-of-the-art fuel-cell density is approximately 1600 Wh/kg. A lightweight HAP, such as a Google balloon weights 28.5 kg, cannot carry heavy long-lasting fuel cells on board. The larger HAP Stratobus can carry up to 450 kg, but it weights already 7 tons leading to high energy consumption for flying. Even if the Stratobus payload capacity is reserved for the HFC, its energy would quickly run out within a few days.

Based on this observation, we believe that \emph{long-duration flights should consider solar energy as the principal energy source}. In this case, the power consumption of a HAP with payload must not exceed the daily harvested solar energy. Let the daily harvested solar energy be $\Psolar$; then,
\begin{equation}
\left(
\begin{aligned}
& [\mPlatform + (\nsFSOi+ \niFSOi) \mFSO] \Pflyunit + \Ppat\\
 &+ \nsFSOi (\PtxFSO+ \Pheat)  + \Pngang  \niFSOi 
  \end{aligned} \right) \leq \frac{\Psolar}{24}
 \label{eq:constraint-solar}
 \end{equation}

According to Figure \ref{fig:energy-beam}, solar energy provision does not need to be very large. A solar energy level between the minimum harvested in York and Enugu allows a 500 kg HAP to carry at least 6 \sFSO~transceivers. A HAP can carry hundreds FSO transceivers with more than 125 kWh solar energy. Therefore, it is realistic to rely on the solar energy. Hereafter, we consider that HAPs solely use solar energy. 

Despite self-sufficient solar energy, HAPs still need to be lowered periodically for maintenance, for example, after one year in the case of Stratobus \cite{Stratobus}. Let us denote the maintenance cycle as a constant $\Dm$. Then 
\begin{equation}
d_i =\Dm, \hspace{1cm}    \forall i \in 1..K
\label{eq:di-fix}
\end{equation}

\section{Optimal \msupp~configuration}
\label{sec:opt1}
 Using multiple \sFSO~transceivers increases the expense of FSO transceivers, although it can reduce the expense of HAPs. This section aims to determine the \msuppconf~that minimizes the HAP network cost defined in \eqref{cost}. We assume that all HAPs use identical \msuppconf s, that is, identical principal beam width $\alpha$, supplementary beam width $\beta$ and number of supplementary \sFSO~transceivers $m$.

Let us now consider the dependence of the HAP network cost on \msuppconf. As each HAP has $m$ supplementary \sFSO~transceivers and uses only solar energy, the cost \eqref{cost} becomes
\begin{equation}
	\Cost=K \CdayH + (Km + \sum_{i=1}^K \niFSOi) \CdayF + \frac{K\Ccharge}{\Dm} \nonumber
\end{equation}

$\Cost$ is a function of $K, m$ and $\niFSOi$. $K$ depends on the coverage radius $\Rext(\alpha, m, \beta)$ of the \msuppconf. $\niFSOi$, as the number of inter-HAP links of HAP $i$, depends on the traffic demand set $\M$. Hence, $\Cost$ depends on \msuppconf~ and $\M$. It is difficult to determine the optimal \msuppconf~without considering $\M$. To relax the dependance on $\M$, we estimate $\Cost$ by a function that depends solely on \msuppconf, that is, tuple $(\alpha, m, \beta)$; then try to find an instance $(\alpha, m, \beta)$ minimizing the estimated cost in expecting that the instance also drives the real cost to a minimum.

\subsection{Cost estimation}
\label{sec:cost-estim}
\begin{figure}[tbh]
\centering
\includegraphics[width=0.45 \textwidth]{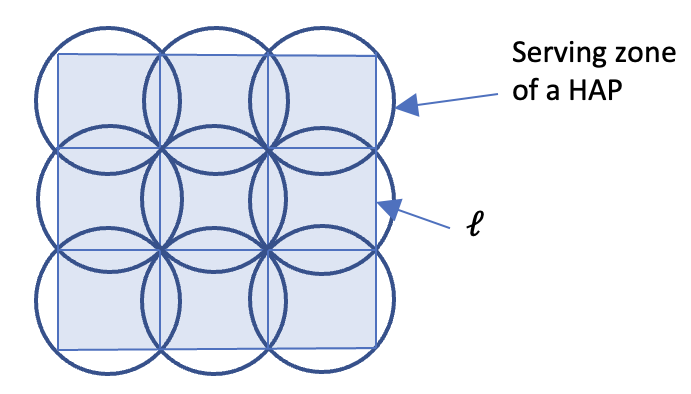}
\caption{A ground area is divided into grid of square cells; each cell is circumscribed by a circle representing a serving zone of a HAP.}
\label{fig:grid}
\end{figure}

First, we estimate the number of HAPs $K$. Samples of the estimation are datasets with uniformly distributed ground nodes. Let $S$ be the surface of the ground area containing those nodes, and $\W$ the number of wavelengths in the WDM technique. We divide the ground zone $S$ into a grid of square cells of size $\ell \times \ell$, each one will be covered by a HAP (see Figure \ref{fig:grid}). To be served by a HAP, a cell must satisfy the following two conditions:
\begin{enumerate} 
\item A cell can contain at most $\W$ ground nodes because a HAP can use at most $\W$ wavelengths to serve ground nodes. Owing to the uniform distribution of ground nodes, we have 
$$\frac{\ell^2}{S}|\NFSO| \leq \W $$
\item A cell must be contained inside by a circle radius equivalent to the extended radius $\Rext$ of a HAP
$$\ell \leq \sqrt{2} \Rext$$
\end{enumerate}
The maximum number of HAPs required to cover region $S$ is the number of cells. Let this number be $\KO$, then,
\begin{equation}
    \KO=\frac{S}{\ell^2}=\lceil \max{\{\frac{|\NFSO|}{\W} , \frac{S}{2 \Rext^2 }\}} \rceil
    \label{eq:K}
\end{equation}
Hence, $\KO$ is an overestimation of the number of HAPs.

Next, we estimate the value of $\niFSOi$. Let $\V$ be the maximum number of inter-HAP links that a HAP may have. Then  
$$\niFSOi \leq \V, \forall i.$$
Finally, $\Cost$ can be overestimated as:
\begin{equation}
	 \CostO =  \KO \left(\CdayH + (m+ \V+ 1)\CdayF + \frac{\Ccharge}{\Dm}\right)
	 \label{eq:cost_over}
\end{equation}

$\CostO$ is a function of $\Rext(\alpha, m, \beta)$ and $m$ while $\V$ is a parameter of the estimator. The estimation is more precise when $\V$ is set close to the actual number of inter-HAP links of a HAP, and coarser otherwise. 

\subsection{Algorithms finding optimal configuration}

Given $\alpha$ and $m$, a larger $\beta$ results in a larger $\Rext$, and thus a smaller $\KO$ and $\CostO$. Therefore, $\beta$ should be set to the largest value according to \eqref{eq:beta-max} for a given $\alpha$ and $m$. It is worth noting that the value of $\beta$ does not affect the solar energy consumption because the laser power $\PtxFSO$ is small and is considered constant. Determining the optimal configuration becomes finding the optimal values of $\alpha$ and $m$.

\begin{algorithm}[tbh]
\caption{Find the optimal \msupp~configuration}
\label{alg:best-m}
\begin{algorithmic}[1]
\Function{Find-optimal-\msupp}{}
\State $\niFSOi \gets \V$ 
\State $cMin \gets \infty$ \Comment{cost min}
\State $\alpha Max \gets$ maximum $\alpha$ by \eqref{eq:alpha-max} 
\For{$\alpha=\alpha Max \ldots 0$}
	\State $mMax \gets$ calculated by \eqref{eq:m-max} \Comment{max $m$}
	\State $mOpt \gets 0$ \Comment{optimal $m$}
	\For{$m=0 \ldots mMax$}
		 \State $\beta \gets$ \Call{Beta-max}{$\alpha$, $m$} \Comment{max $\beta$}
 		\State Calculate $\Rext(\alpha, m, \beta)$ using \eqref{eq:r-extended-1},\eqref{eq:axt},\eqref{eq:tan-gamma} \eqref{eq:tan-phi}
		\State Calculate $\CostO(\alpha, m, \beta)$ using \eqref{eq:cost_over}
		 \If{$\CostO < cmin$}
			\State $cmin \gets \CostO$
			\State $\alpha Opt \gets \alpha$ \Comment{optimal $\alpha$}
			\State $mOpt \gets m$ \Comment{optimal $m$}
			\State $\beta Opt \gets \beta$ \Comment{optimal $\beta$}
 		\EndIf
	\EndFor
\EndFor
\State \Return $\alpha Opt, mOpt, \beta Opt$
\EndFunction
\end{algorithmic}
\end{algorithm}
\begin{algorithm}[tbh]
\caption{Find the maximum $\beta$ given $\alpha,m$}
\label{alg:beta-max}
\begin{algorithmic}[1]
\Function{Beta-max}{$\alpha$, m}
\For{$\beta=0 \ldots 180$}
 \State Calculate $\Rext (\alpha, m, \beta)$ using \eqref{eq:r-extended-1},\eqref{eq:axt},\eqref{eq:tan-gamma} \eqref{eq:tan-phi}
 \State Calculate $L_J$ using \eqref{eq:Ljbeta-1}
 \State Calculate $P_{J}^{rx}$ using \eqref{eq:Prx-beta}
  \If{$P_{J}^{rx} $ <$\Prx$}		\Comment{Looking for the first $\beta$ violate constraint \eqref{eq:beta-max}}
	\State \Return $\beta$-1	\Comment{the previous trial $\beta$ was the maximum}	
 \EndIf
\EndFor
\EndFunction
\end{algorithmic}
\end{algorithm}

Following an exhaustive search approach, we examine all possible values of $\alpha$ and $m$ to seek for the pair that minimizes $\CostO$ in \eqref{eq:cost_over}. The search range of $\alpha$ is from $0\degree$ to the maximum value set by constraint \eqref{eq:alpha-max}. The number of supplementary \sFSO~transceivers $m$ is also limited. Indeed, since the number of inter-HAP links of a HAP can go up to $\V$ as set in Section \ref{sec:cost-estim}, and $\nsFSOi=m+1, \forall i$, then from the energy constraint \eqref{eq:constraint-solar}, we deduce the upper bound for $m$:
\begin{equation}
    m \leq \frac{\frac{\Psolar}{24} - (\V \mFSO \Pflyunit + \V \Pngang + \mPlatform \Pflyunit + \Ppat)}{\mFSO  \Pflyunit + \Pheat + \PtxFSO} -1
    \label{eq:m-max}
\end{equation}

Algorithm \ref{alg:best-m} implements the exhaustive search idea. First, two nested loops scan all possible values of $\alpha$ satisfying constraint \eqref{eq:alpha-max} and all possible values of $m$ satisfying \eqref{eq:m-max} to find the pair that minimizes $\CostO$ in \eqref{eq:cost_over}. For each pair $(\alpha, m)$, the largest value of $\beta$ according to constraint \eqref{eq:beta-max} is selected using Algorithm \ref{alg:beta-max}. 
The optimal \msuppconf~is reported by the algorithms as $(\alpha Opt, mOpt, \beta Opt)$.
 
Algorithm \ref{alg:beta-max} finds the maximum $\beta$ that satisfies constraint \eqref{eq:beta-max} for a given pair of $(\alpha, m)$ by testing the possible values of $\beta$ increasingly from 0 until the received power $P_{J}^{rx}$ at the border of the extended coverage area reaches the required received power $\Prx$. The received power $P_{J}^{rx}$ is calculated using the set of equations \eqref{eq:Prx-beta},  \eqref{eq:Ljbeta-1}, \eqref{eq:r-extended-1},\eqref{eq:axt},\eqref{eq:tan-gamma}, and \eqref{eq:tan-phi}. 

In the implementation of both algorithms, $\alpha$ and $\beta$ step by $1 \degree$ after each iteration. Finer stepping allows obtaining more accurate results. However, even with $1 \degree$ stepping, the variation in the optimal $\Rext$ is only a few hundred meters, which is negligible in comparison to the absolute value of $\Rext$ which is in the range of 6-30 kilometers. 

The complexity of Algorithm \ref{alg:best-m} is $O(m)$ because $\alpha \leq \pi$. The complexity of Algorithm \ref{alg:beta-max} is constant because $\beta \leq \pi$.


\section{Design HAP network topology}
\label{sec:algo}

This section presents the HAP network design using the optimal configuration identified above. Let denote $\Linter$ as the number of inter-HAP links. Since $\sum_{i=1}^K \niFSOi$ is the total number of inter-HAP FSO transceivers, it is equal to $2 \Linter$. The network cost becomes:
\begin{equation}
	 \Cost = K \CdayH + ( K(m+1) + 2\Linter)  \CdayF + K \frac{\Ccharge}{\Dm} \nonumber
\end{equation}
and is equivalent to
\begin{equation}
	 \Cost =  K \left(\CdayH + (m+1) \CdayF + \frac{\Ccharge}{\Dm}\right) +  2 \Linter\CdayF
	 \label{eq:cost_solar}
\end{equation}

The cost is proportional to the number of HAPs $K$ and the number of inter-HAP links $\Linter$. We consider that the daily amortization cost of a HAP is much greater than that of an FSO transceiver; thus, the coefficient of $K$ is much greater than the coefficient of $\Linter$ in $\Cost$. Consequently,  $K$ should be prioritized to minimize over $\Linter$. Therefore, the topology design is broken into following two steps:
\begin{itemize}
\item[i)] ground nodes are clustered into equal radius circles that will become serving zones of HAPs in such a way that the number of clusters is the smallest for minimizing $K$; 
\item[ii)] corresponding HAPs are located at the centers of clusters but at an elevation of 20 km and are interconnected by the fewest number of inter-HAP links, $\Linter$.
\end{itemize}
A HAP network topology design algorithm was proposed in \cite{osnpaper} following these two steps. In this algorithm, the clustering radius was not determined but was left as an input of the algorithm. In the current study, we set the clustering radius as the extended coverage radius $\Rext$ of the optimal \msupp~configuration to drive towards a HAP network with minimal $\Cost$.
\begin{figure}[tbh]
\centering
\includegraphics[width=0.4\textwidth]{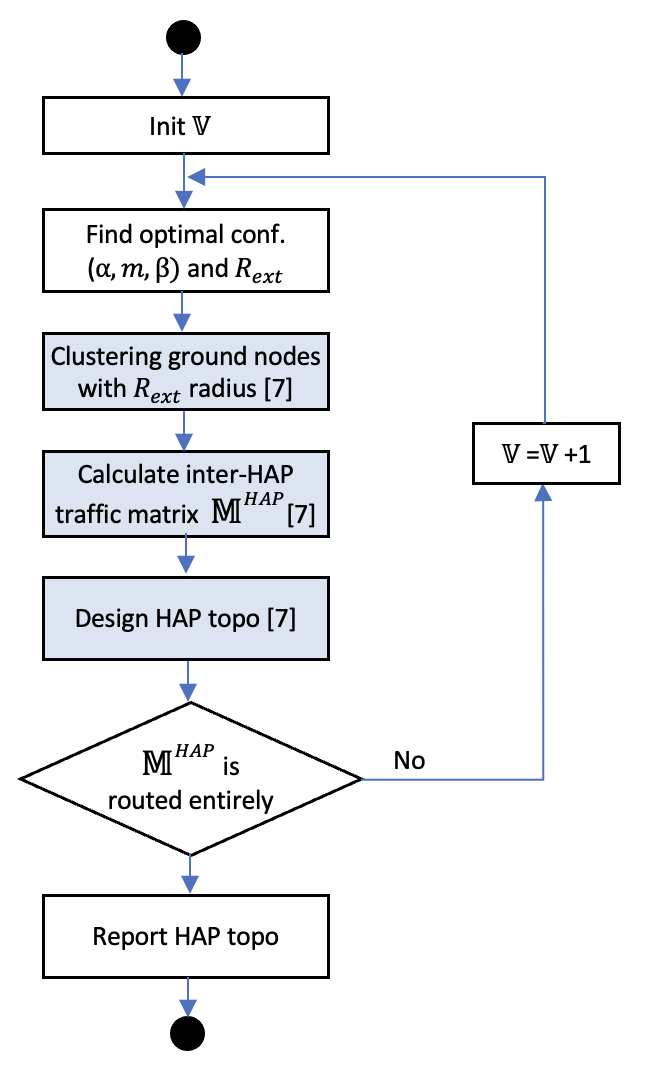}
\caption{HAP network design flowchart.}
\label{fig:topo-design-flowchart}
\end{figure}
The main steps of the HAP network design process are presented in Figure \ref{fig:topo-design-flowchart}, where the steps taken from \cite{osnpaper} are shown in color. The process is explained as follows:
\begin{itemize}
	\item  Initialize $\V$, the maximum number of inter-HAP links of a HAP, by a constant.
	\item Calculate the optimal \msupp~configuration using Algorithm \ref{alg:best-m}, and set the clustering radius as its $\Rext$.
	\item Apply the clustering algorithm proposed in \cite{osnpaper} to distribute ground nodes into clusters of radius $\Rext$ while keeping the number of ground nodes in each cluster under $\W$. Each cluster becomes a serving zone of a HAP. The HAP is located at the center of the cluster but at an elevation of 20 km.
	\item Bandwidth demands between ground nodes belonging to different serving zones are bundled into lightpaths between corresponding HAPs, creating the inter-HAP traffic matrix $\MHAP$. 
	\item Apply HAP topology design algorithm proposed in \cite{osnpaper} to build the HAP topology. The algorithm begins with an empty topology. It finds a route for each lightpath demand of $\MHAP$ from a full-mesh graph linking all HAPs within communication distance limit $\LHH$. Each time a lightpath uses an inter-HAP link that has not yet been included in the current HAP topology, the link is incorporated into the topology. The link in the topology is prioritized for use in building routes for the next lightpath demands. 
	\item Once all lightpath demands in $\MHAP$ are routed, the final topology is achieved. Otherwise, routing may fail due to the low connectivity between HAPs. In this case, $\V$ is increased by one, and the process is repeated until all lightpath demands in $\MHAP$ are routed.
	 \end{itemize}

\section{Simulation results}
\label{sec:experiment}
The algorithms for finding the optimal \msupp~configuration were implemented and integrated with the topology designed algorithm described in Section \ref{sec:algo}. We performed simulations with practical parameters and evaluated the efficiency of \msupp~configuration compared to the single \sFSO~transceiver configuration.

\subsection{Parameter values}
\label{sec:param}

The simulation parameters are listed in Table \ref{tab:param}. The values of these parameters were chosen according to experiments reported in the literature. This subsection explains the choices of the parameter values.

\emph{Cost-related parameters:} The cost-related parameters are set such that the daily amortization cost of a HAP is significantly greater than that of an FSO transceiver, and the one-time maintenance cost is significantly higher than the daily amortization cost of a HAP. The maintenance cycle of a HAP is set as $\Dm=1$~year according to published information on Stratobus \cite{Stratobus}.

\emph{Energy-related parameters:} 
\begin{itemize}
    \item $\Psolar$ - daily harvested solar energy. We considered daily solar energy levels between the minimum daily solar energy values in York and Enugu reported in  \cite{solar-energyHAP2020}, which were 42 kWh and 290 kWh, respectively.
    \item $\Pflyunit$ - power consumed by the avionic part of a HAP to carry a unit of mass. Although the power-to-mass ratio can be estimated as 6 W/kg according to \cite{solar-energyHAP2020}, the published power rates of real systems are smaller. For aerodynamic systems such as Zephir-S, Zephir-T \cite{Zephir}, and Phasa-35 \cite{Phasa}, $\Pflyunit$ varies from 2.68 -3.04 W/kg. Indeed, Zephir-S weighs 80 kg (75 kg platform and 5 kg payload) and consumes 243 W, Zephir-T weighs 160 kg (140 kg platform and 20 kg payload) and consumes 429 W, and Phasa-35 weighs 165 kg (150 kg platform and 15 kg payload) and consumes 459 W. Aerostatic systems consume even less power. The Stratobus weighs 7000 kg and consumes 5~kW when it carries a 250 kg payload and 8~kW when it carries 450~kg \cite{Stratobus}. Thus, the power-to-mass ratio of Stratobus is between 0.69 and 1.07 W/kg only. Therefore, in this simulation $\Pflyunit$ was set to 2~W/kg.
    \item $\Ppat$ - power consumed by a PAT system; it was set to 15 W according to \cite{Optical-HAP-2010}. 
    \item $\Pheat$ - power consumed for heating, cooling, and management; it was set to 20 W according to \cite{LoonProject}.
    \item $\Pngang$ - power consumed by an inter-HAP FSO transceiver; it was set to 35.1 W including laser power, $\Pheat$ and $\Ppat$.
	\end{itemize}

\emph{Inter-HAP link parameters:} These parameters were set to values similar to those provided in the Loon project \cite{LoonProject}.

\emph{HAP-ground FSO link parameters:} The attenuation coefficient of an FSO link between a HAP and a ground node is set identical to that of inter-HAP links. The required received power $\Prx$ at a ground node was set according to \cite{LoonProject}. The aperture radius $\Rrx$ of a ground FSO receiver was set according to the commercial FSO transceiver SONABeam \cite{fSona}.

\emph{Other parameters:}
	\begin{itemize}
		\item $\delta$ - BER threshold for inter-HAP links and lightpaths. We set $\delta=10^{-3}$ because errors with that BER can be corrected using current Forward Error Correction (FEC) techniques.
		\item $\LHH$ - the maximum allowable distance between two HAPs such that the BER of an inter-HAP link is less than $\delta=10^{-3}$. Using the inter-HAP FSO link parameters listed in Table \ref{tab:param}, the calculation yielded $\LHH=88$~km.
		\item $\mPlatform$ - platform mass; it varies significantly from one design to another. The Loon balloon weighs just 28.5 kg while the Stratobus weighs 7000 kg. With $\Pflyunit=2$~W/kg, a HAP weighing more than 7000 kg already consumes 326 kWh/day to carry itself, which is more than the maximum harvested solar energy, leading to no remaining energy to carry FSO transceivers. Therefore, $\mPlatform=500$ kg was used in the simulations.
		\item $\mFSO$ - mass of an FSO transceiver on HAPs. It was set according to the FSO transceiver used in the Loon project, which weighs 6.3 kg \cite{LoonProject}. This value is consistent with the weights between 8 and 10~kg of commercial terrestrial SONABeam FSO transceivers\cite{fSona}. 
		\item $\W$ - the number of wavelengths per FSO link. It was set to 40 or 80 according to the current WDM technique.
	\end{itemize}

The test dataset contained 19 test cases, each with 400 -- 2800 ground nodes. The ground FSO node locations were randomly generated on a square surface of $100 \times 100$ km, which is the size of a large metropolis. The test cases had different numbers of ground nodes, reflecting different ground node densities. The traffic requirement $\M$ contained demands randomly generated between ground FSO nodes such that the total incoming or outgoing traffic of a ground FSO node did not exceed 1 Gbps, which is the capacity of a single wavelength. 

Initially, $\V$ was set to $10$. The optimal multiple \sFSO~transceiver configuration $(\alpha, m, \beta)$ was calculated using Algorithms \ref{alg:best-m} and \ref{alg:beta-max}. The extended radius $\Rext$ of the optimal configurations was calculated using \eqref{eq:r-extended-1} and was then used as the clustering radius in the HAP topology design step. 

With $\Psolar =42$ kWh and $\W=40$,  $\V$ must be increased to 12 to get all demands in $\MHAP$ routed successfully for all test cases. With all other $\Psolar$ and $\W$ values, the topology design algorithm successfully routed all demands in $\MHAP$ for all test cases right with initial $\V=10$. 

\begin{figure}[tbh]
     \includegraphics[width=0.45 \textwidth]{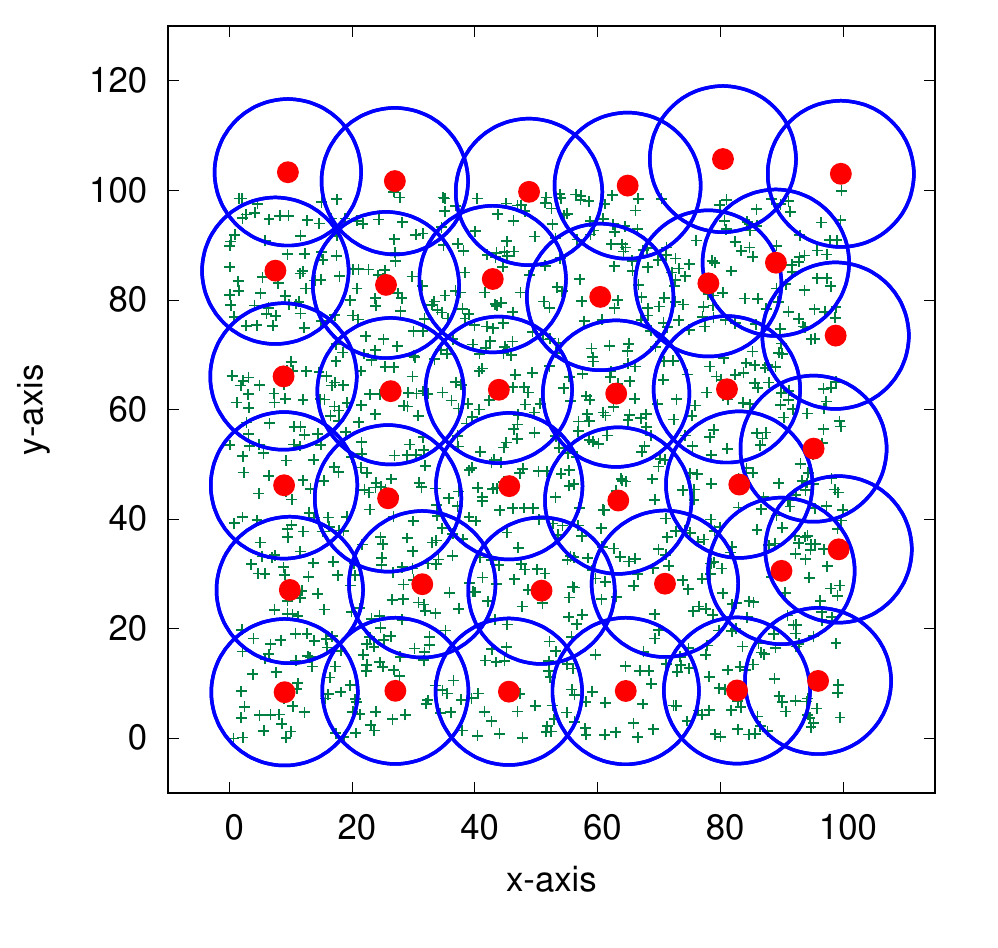}
    \caption{Footprints of HAPs with \msupp~configuration obtained from the topology design  for a test case of 1005 ground FSO nodes when $\Psolar=75$~kwh, $\W=80$. A circle represents an extended coverage area of a HAP. Small points inside the circle are ground nodes and the dot at the center of the circle is the projected location of its serving HAP on the ground. }
    \label{fig:ex-cluster}
\end{figure}

Figure \ref{fig:ex-cluster} illustrates the HAP locations and their footprints calculated using the proposed algorithms for a test case of 1005 ground FSO nodes, $\Psolar=75$~kWh, and $\W=80$. 

\begin{table*}[tbh]
\begin{tabular}{|l|r|r|r|r|r|r|r|r|r|r|r|r|r|r|r|r|r|r|r|r|r}
\hline
        & \multicolumn{10}{|c|}{$\Psolar=42$ kWh} & \multicolumn{10}{|c|}{$\Psolar=50 \sim 290$ kWh}\\ \cline {2-21}
	&\multicolumn{5}{|c|}{$\W=40, \V=12$} & \multicolumn{5}{|c|}{$\W=80, \V=10$} & \multicolumn{5}{|c|}{$\W=40, \V=10$} & \multicolumn{5}{|c|}{$\W=80, \V=10$} \\ \cline {2-21}
$|\NFSO|$ & $\alpha$& $m$ & $\beta$ & $\Rext$ &Cost & $\alpha$ & $m$ & $\beta$ & $\Rext$ &Cost & $\alpha$& $m$ & $\beta$ & $\Rext$ &Cost & $\alpha$& $m$ & $\beta$ & $\Rext$ & Cost\\
\tiny{(1)}&\tiny{(2)}&\tiny{(3)}&\tiny{(4)}&\tiny{(5)}&\tiny{(6)}&\tiny{(7)}&\tiny{(8)}&\tiny{(9)}&\tiny{(10)}&\tiny{(11)}&\tiny{(12)}&\tiny{(13)}&\tiny{(14)}&\tiny{(15)}&\tiny{(16)}&\tiny{(17)}&\tiny{(18)}&\tiny{(19)}&\tiny{(20)}&\tiny{(21)}\\
\hline
480	&	37	&	0	&	-	&	6691	&	15308	&	37	&	0	&	-	&	6691	&	13488	&	37	&	13	&	16	&	11929	&	10010	&	37	&	13	&	16	&	11929	&	9470	\\
588	&	37	&	0	&	-	&	6691	&	16639	&	37	&	0	&	-	&	6691	&	14519	&	37	&	13	&	16	&	11929	&	9304	&	37	&	13	&	16	&	11929	&	8964	\\
763	&	37	&	0	&	-	&	6691	&	18495	&	37	&	0	&	-	&	6691	&	15815	&	37	&	13	&	16	&	11929	&	9990	&	37	&	13	&	16	&	11929	&	9510	\\
854	&	37	&	0	&	-	&	6691	&	19141	&	37	&	0	&	-	&	6691	&	16341	&	37	&	13	&	16	&	11929	&	10493	&	37	&	13	&	16	&	11929	&	9933	\\
998	&	37	&	0	&	-	&	6691	&	19101	&	37	&	0	&	-	&	6691	&	16701	&	37	&	13	&	16	&	11929	&	10855	&	37	&	13	&	16	&	11929	&	10215	\\
1005	&	37	&	0	&	-	&	6691	&	19068	&	37	&	0	&	-	&	6691	&	16308	&	37	&	13	&	16	&	11929	&	11138	&	37	&	13	&	16	&	11929	&	10478	\\
1150	&	37	&	0	&	-	&	6691	&	19666	&	37	&	0	&	-	&	6691	&	16926	&	37	&	13	&	16	&	11929	&	10915	&	37	&	13	&	16	&	11929	&	10275	\\
1345	&	37	&	0	&	-	&	6691	&	20644	&	37	&	0	&	-	&	6691	&	17564	&	37	&	13	&	16	&	11929	&	11741	&	37	&	13	&	16	&	11929	&	10395	\\
1477	&	37	&	0	&	-	&	6691	&	20752	&	37	&	0	&	-	&	6691	&	17612	&	37	&	12	&	16	&	11539	&	13115	&	37	&	13	&	16	&	11929	&	10335	\\
1523	&	37	&	0	&	-	&	6691	&	21895	&	37	&	0	&	-	&	6691	&	18595	&	37	&	12	&	16	&	11539	&	14053	&	37	&	13	&	16	&	11929	&	10375	\\
1675	&	37	&	0	&	-	&	6691	&	21735	&	37	&	0	&	-	&	6691	&	18535	&	37	&	11	&	16	&	11042	&	14128	&	37	&	13	&	16	&	11929	&	10495	\\
1736	&	37	&	0	&	-	&	6691	&	22461	&	37	&	0	&	-	&	6691	&	19301	&	37	&	11	&	16	&	11042	&	14874	&	37	&	13	&	16	&	11929	&	10455	\\
1911	&	37	&	0	&	-	&	6691	&	22481	&	37	&	0	&	-	&	6691	&	19021	&	37	&	10	&	16	&	10395	&	14869	&	37	&	13	&	16	&	11929	&	10495	\\
2009	&	37	&	0	&	-	&	6691	&	22641	&	37	&	0	&	-	&	6691	&	19321	&	37	&	10	&	16	&	10395	&	15575	&	37	&	13	&	16	&	11929	&	10595	\\
2135	&	37	&	0	&	-	&	6691	&	22761	&	37	&	0	&	-	&	6691	&	19221	&	37	&	10	&	16	&	10395	&	16493	&	37	&	13	&	16	&	11929	&	10575	\\
2304	&	37	&	0	&	-	&	6691	&	22881	&	37	&	0	&	-	&	6691	&	19301	&	37	&	9	&	16	&	9524	&	18192	&	37	&	13	&	16	&	11929	&	10655	\\
2325	&	37	&	0	&	-	&	6691	&	22368	&	37	&	0	&	-	&	6691	&	18948	&	37	&	9	&	16	&	9524	&	18555	&	37	&	13	&	16	&	11929	&	10675	\\
2491	&	37	&	0	&	-	&	6691	&	22761	&	37	&	0	&	-	&	6691	&	19401	&	37	&	8	&	16	&	8946	&	18660	&	37	&	13	&	16	&	11929	&	10655	\\
2753	&	37	&	0	&	-	&	6691	&	23346	&	37	&	0	&	-	&	6691	&	19926	&	37	&	8	&	16	&	8946	&	20284	&	37	&	13	&	16	&	11929	&	11178	\\
\hline
\end{tabular}
\caption{Optimal configurations and costs of all test cases with $\Rrx=2$ m.} 
\label{tab:best-radius}
\end{table*}

\begin{table}[tbh]
\centering
\begin{tabular}{|r|r|r|}
\hline
Receiver aperture  & Maximum beam  & Maximum \\ 
 radius $\Rrx$ (m) &  width $\alpha_{max}$ & coverage radius (m) \\ 
\hline
2 & 37 \degree & 6691\\
4 & 67 \degree & 13237\\ \hline
\end{tabular}
\caption{Maximum beam width and coverage radius of single \sFSO~transceiver configuration.}
\label{tab:max-alpha}
\end{table}
\begin{table} [tbh]
\centering
\begin{tabular}{|r|r|r|r|r|}
\hline
$\Psolar$ &  		& \multicolumn{2}{|c|}{Max $\Rext$ (m)} \\ \cline{3-4}
(kWh)	&  Max $m$ & $\Rrx=2$ (m) &  $\Rrx=4$ (m) \\
\hline
42 & 6 & 6691 & 13237 \\ \hline
50 &16 & 12174 &25582\\ \hline
75 &47 & 13559 & 28403\\ \hline
100 &78 & 13678 &  28845 \\ \hline
125 &109 & 13711 &28969 \\ \hline
150 &140 & 13724 & 29020 \\ \hline
175  &171 & 13731 &29047 \\ \hline
200 &202 & 13735 & 29062 \\ \hline
225 & 233 & 13738  &29071 \\ \hline
250 &264 & 13739  &29077 \\ \hline
275 &295 & 13740 &29082 \\ \hline
290 &314 &13741 &29084 \\
\hline
\end{tabular}
\caption{Maximum extended coverage radius of \msupp~configuration when $\V=10$.}
\label{tab:beamwidth}
\end{table}
\subsection{\msupp~configuration versus single serving FSO transceiver configuration} 

Table \ref{tab:max-alpha} lists the maximum beam width $\alpha_{max}$  according to \eqref{eq:alpha-max} and the maximum ground coverage radius of the single \sFSO~transceiver configuration when the receiver aperture radius was varied. Table \ref{tab:beamwidth}  lists the extended coverage radius of the maximum \msupp~configuration for different solar energy levels and receiver aperture radii. The maximum \msupp~configuration was obtained using the largest principal beam $\alpha_{max}$, largest $m$ according to \eqref{eq:m-max}, and largest $\beta$ according to \eqref{eq:beta-max}, given $\alpha_{max}$ and $m$.  The coverage radius of the maximum \msupp~configuration was extended approximately twice in comparison with that of single FSO transceiver configuration, except for $\Psolar=42$kWh.  When solar energy level increased, the maximum $m$ increased; thus, the extended coverage radius increased. However, when $m$ was already large, the extention increased slowly with $m$. Additionally, the maximum extended coverage was much larger when $\Rrx=4$ than $\Rrx=2$m because a receiver can accept weaker signals with larger apertures. 

To compare the network costs incurred by the two configurations, we examined the detailed results in Table \ref{tab:best-radius}. The table lists the optimal \msupp~configurations and network costs. When $\Psolar=42$kWh, the optimal number of supplementary \sFSO~transceivers is $m=0$; thus, the configuration uses a single \sFSO~transceiver. Therefore, these cases were used as references for single \sFSO~transceiver configuration. When $\Psolar > 50$kWh, all optimal configurations were truly \msupp, and the results were identical for all solar energy levels. The numbers indicate that \msupp~configuration offered significantly lower costs (listed in columns $16^{th}$ and $21^{th}$) than those of single \sFSO~transceiver configuration (listed in columns $6^{th}$ and $11^{th}$) for the same test cases and number of wavelengths $\W$. The costs resulting from \msupp~configuration were as low as 54--87\% of those resulting from single \sFSO~transceiver configuration.  \emph{These numbers confirm that when there is sufficient solar energy, \msupp~configuration is definitively a better choice than single \sFSO~configuration.}

\subsection{Factors impact optimal \msupp~configuration}

Comparing the values of the optimal extended coverage radius in Table \ref{tab:best-radius}  and the maximum extended coverage radius in Table \ref{tab:beamwidth}, we can see that the optimal extended coverage radius was generally not the maximum. This is reasonable because the maximum configuration uses an excessive number of supplementary \sFSO~transceivers. 

Low solar energy may render \msupp~configuration impossible. Indeed, $\Psolar=42$ kWh could afford maximally 6 supplementary \sFSO~transceivers (see Table \ref{tab:beamwidth}), which was too few to entirely cover the contour of the principal coverage area. Thus, single \FSO~transceiver configuration was the unique choice.

When the solar energy level exceeds 50 kWh, its exact value does not affect the optimal configuration. The simulation showed that the optimal configurations were identical for all solar energy levels from 50 kWh/day and above. This is explained by the fact that a greater solar energy level allows to accept configurations with large coverage but may be more expensive because of using more supplementary \sFSO~transceivers. As a result, large configurations were not selected as optimal configurations. \emph{In other words, increasing solar energy does not necessarily improve the HAP network cost.}
 
Since the optimal multiple \sFSO~transceiver configurations were identical for all $\Psolar \geq 50$ kWh, all other numerical results related to topology design and routing with these solar energy levels were identical and are presented as single results in subsequent figures.
 
The coverage of the optimal configurations decreased when the ground nodes became denser. Indeed, test cases with large numbers of ground nodes had greater ground node densities, and columns $13^{th}$ and $16^{th}$ of Table \ref{tab:best-radius} shows that the optimal $m$ and $\Rext$ decreased when the density increased. The reason is that, with a greater ground node density, is a small ground region already contains $\W$ ground nodes, which is the maximum serving capacity of a HAP. Therefore, a HAP could serve only a small zone and required only a few supplementary FSO transceivers to cover the zone.

\subsection{Numbers of HAPs and inter-HAP links}
\label{sec:ex:nb-HAP}
\begin{figure}[tbh]
    \centering
     \subfigure[W=40]{\includegraphics[width=0.49\textwidth]{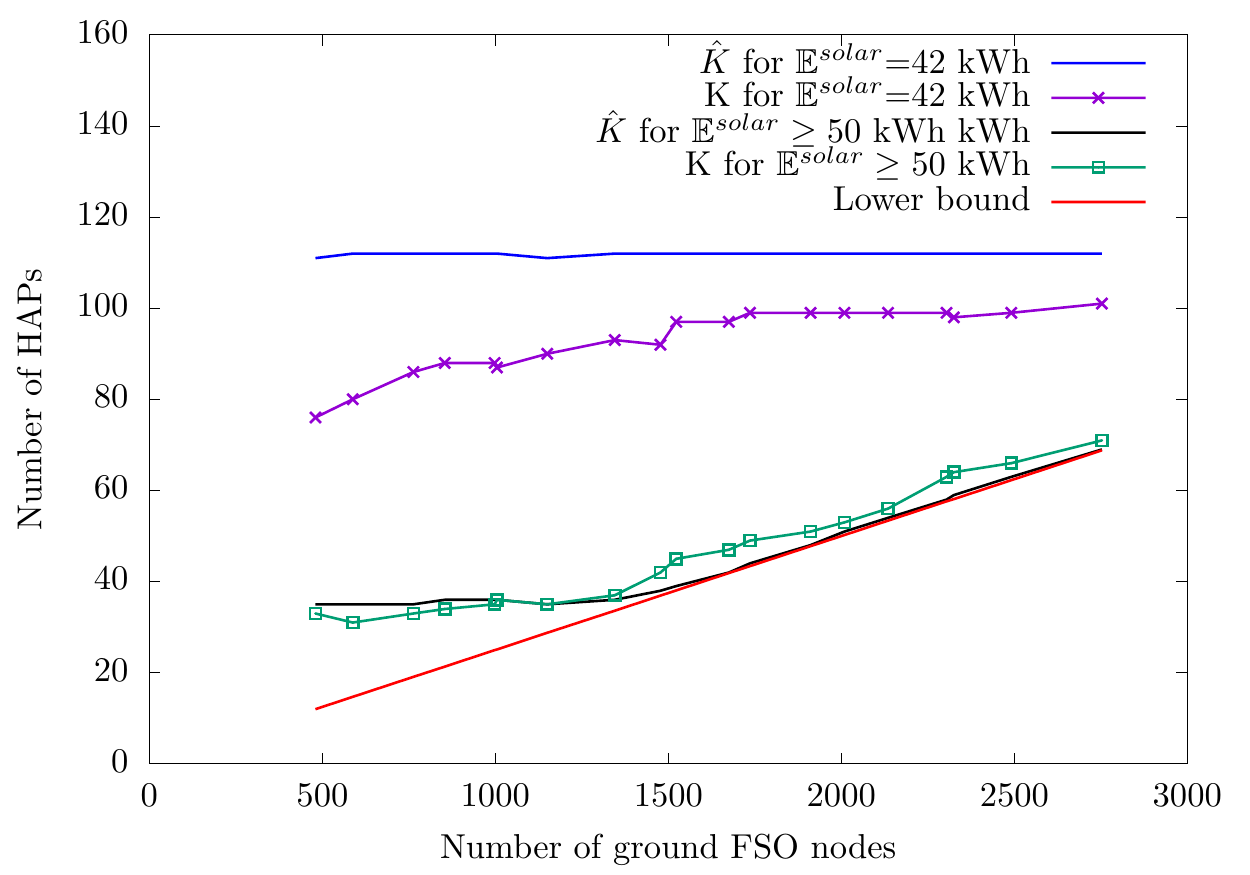}}
     \subfigure[W=80]{\includegraphics[width=0.49\textwidth]{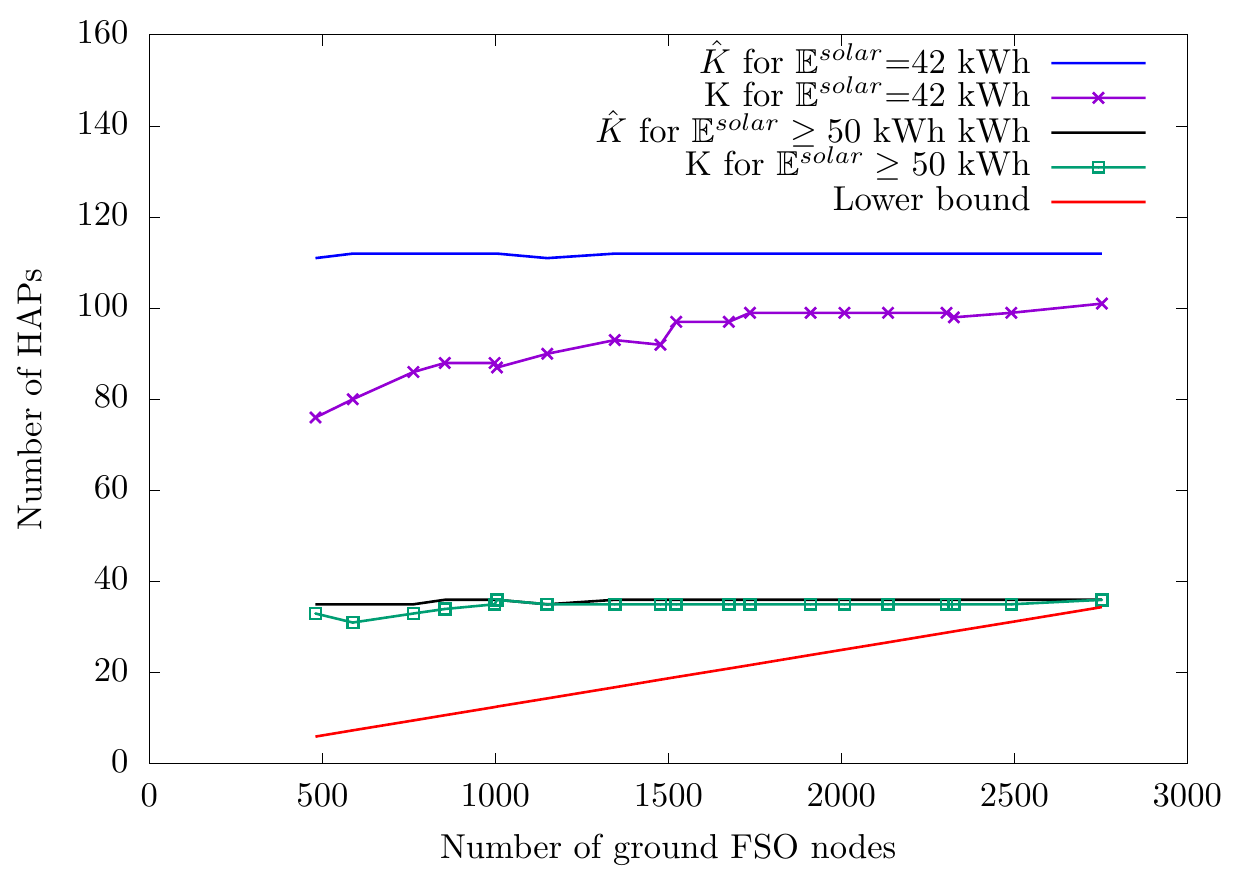}}
    \caption{Number of HAPs and lower bound with (a) $\W=40$ and (b) $\W=80$ in different solar energy levels.}
    \label{fig:nb-HAP-LB}
\end{figure}

\begin{figure}[tbh]
    \centering
    \subfigure[W=40]{\includegraphics[width=0.48\textwidth]{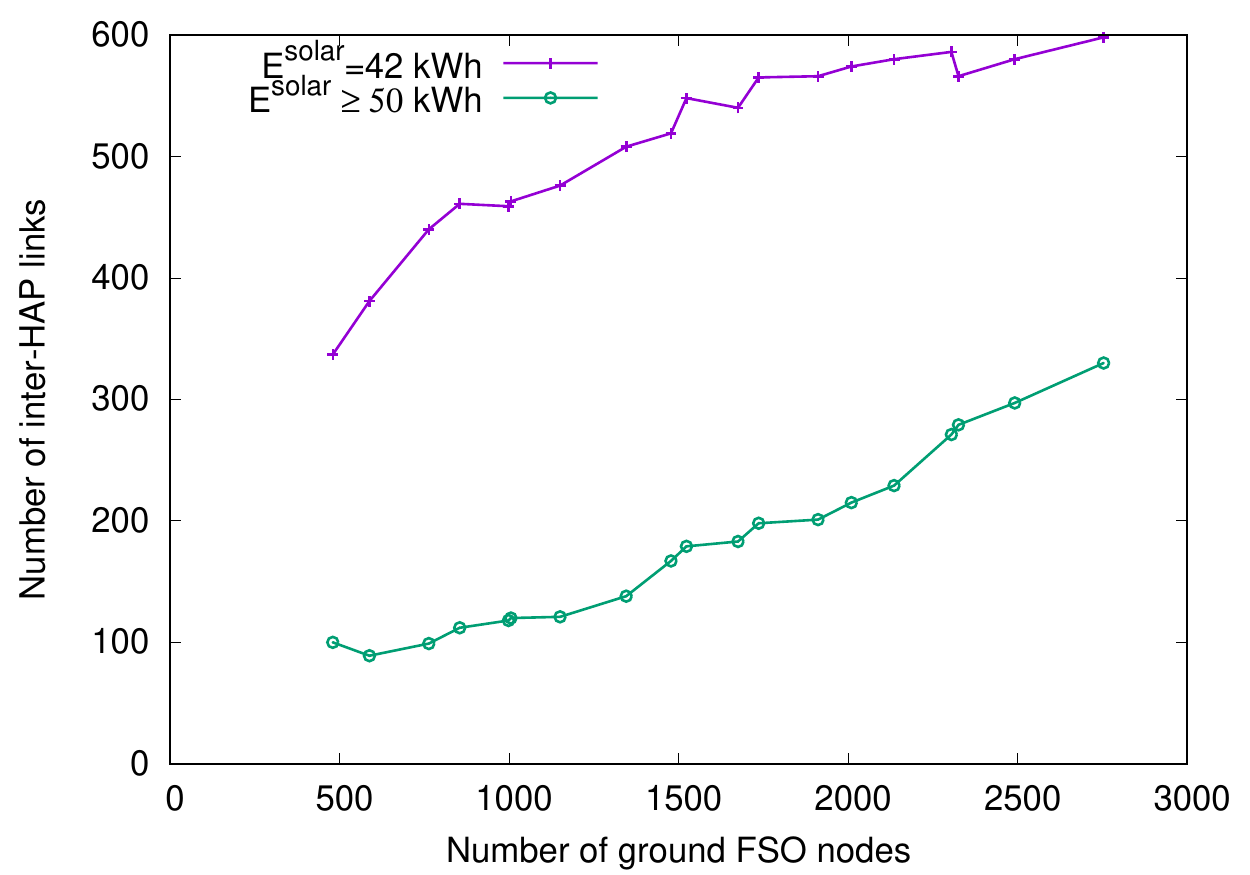}}
    \subfigure[W=80]{\includegraphics[width=0.48\textwidth]{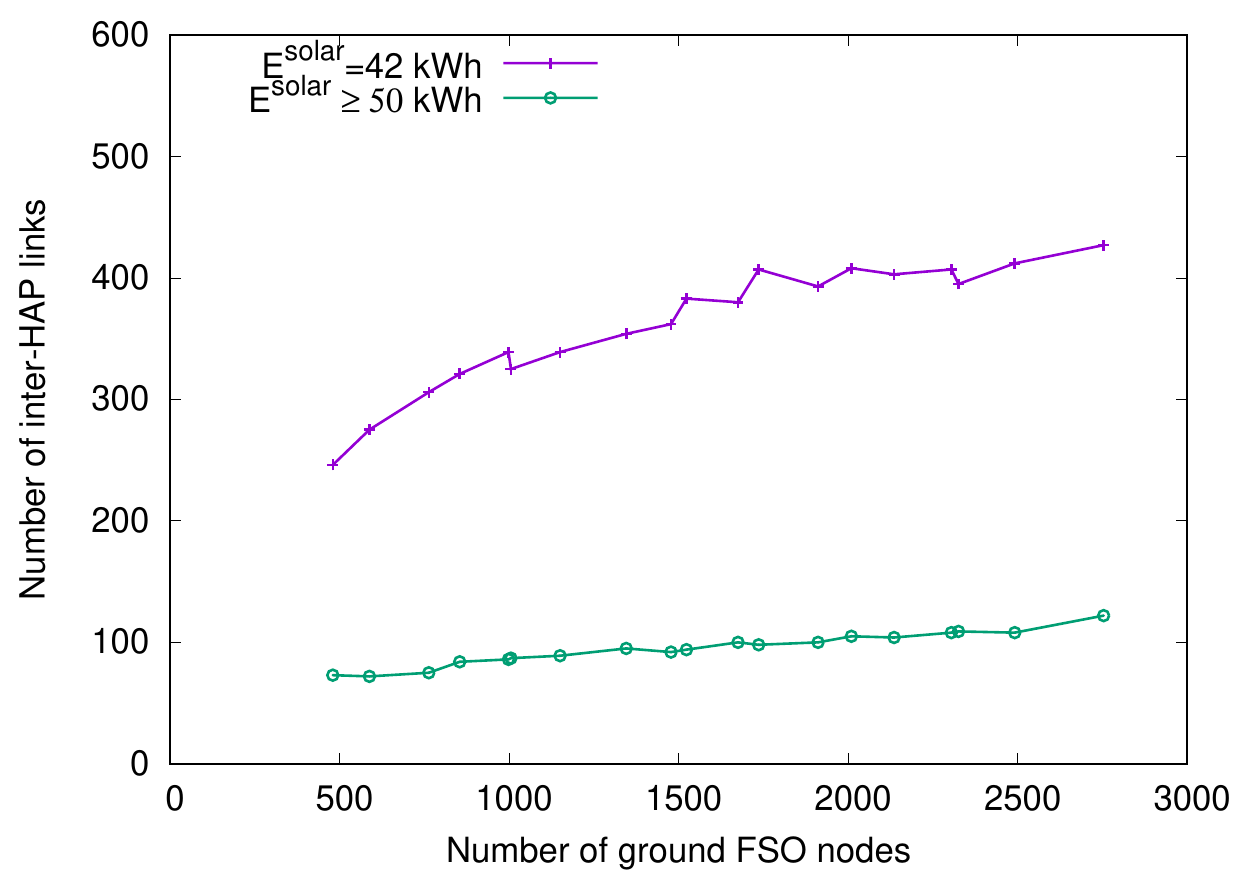}}
    \caption{Number of inter-HAP links when (a) $\W=40$ and (b) $\W=80$ for different solar energy levels.}
    \label{fig:nb-inter-HAP-link-all}
\end{figure}

Since each HAP can serve at most $\W$ ground FSO nodes, a lower bound for the number of HAPs is:
\begin{equation}
    n_{\HAP}^{LB}=\frac{|\N_{\FSO}|}{\W}
\label{eq:lowerbound}
\end{equation}

Figure \ref{fig:nb-HAP-LB} shows the number of HAPs, the estimated number of HAPs $\hat{K}$ and lower bound $ n_{\HAP}^{LB}$ when (a) $\W=40$ and (b) $\W=80$. With $\Psolar \geq 50$~kWh, the actual number of HAPs was almost identical to $\hat{K}$ in both subfigures. Furthermore, when $\W=40$ and $\Psolar \geq 50$~kWh, the number of HAPs approached the lower bound starting from test cases with 1000 ground nodes or above. This implies that the number of HAPs was almost optimal.

Figure \ref{fig:nb-inter-HAP-link-all} presents  the absolute numbers of inter-HAP links. The number of inter-HAP links increased with the number of ground nodes, because the network size and traffic demand increased. The number of inter-HAP links clearly decreased when the wavelength density increased from $\W=40$ to $\W=80$. In other words, denser WDM technique helps reduce the number of inter-HAP FSO transceivers and consequently the network cost.

\msupp~configuration allows reducing significantly both the numbers of HAPs and inter-HAP links. Indeed, according to Figure \ref{fig:nb-HAP-LB}, the number of HAPs was much smaller with  $\Psolar \geq 50$ kWh where \msupp~configuration was used, in comparison with $\Psolar = 42$ kWh, where single \sFSO~configuration was used. A similar phenomenon is observed in Figure \ref{fig:nb-inter-HAP-link-all} for the number of inter-HAP links.


\begin{figure}[tbh]
    \centering
    \subfigure[W=40]{\includegraphics[width=0.47\textwidth]{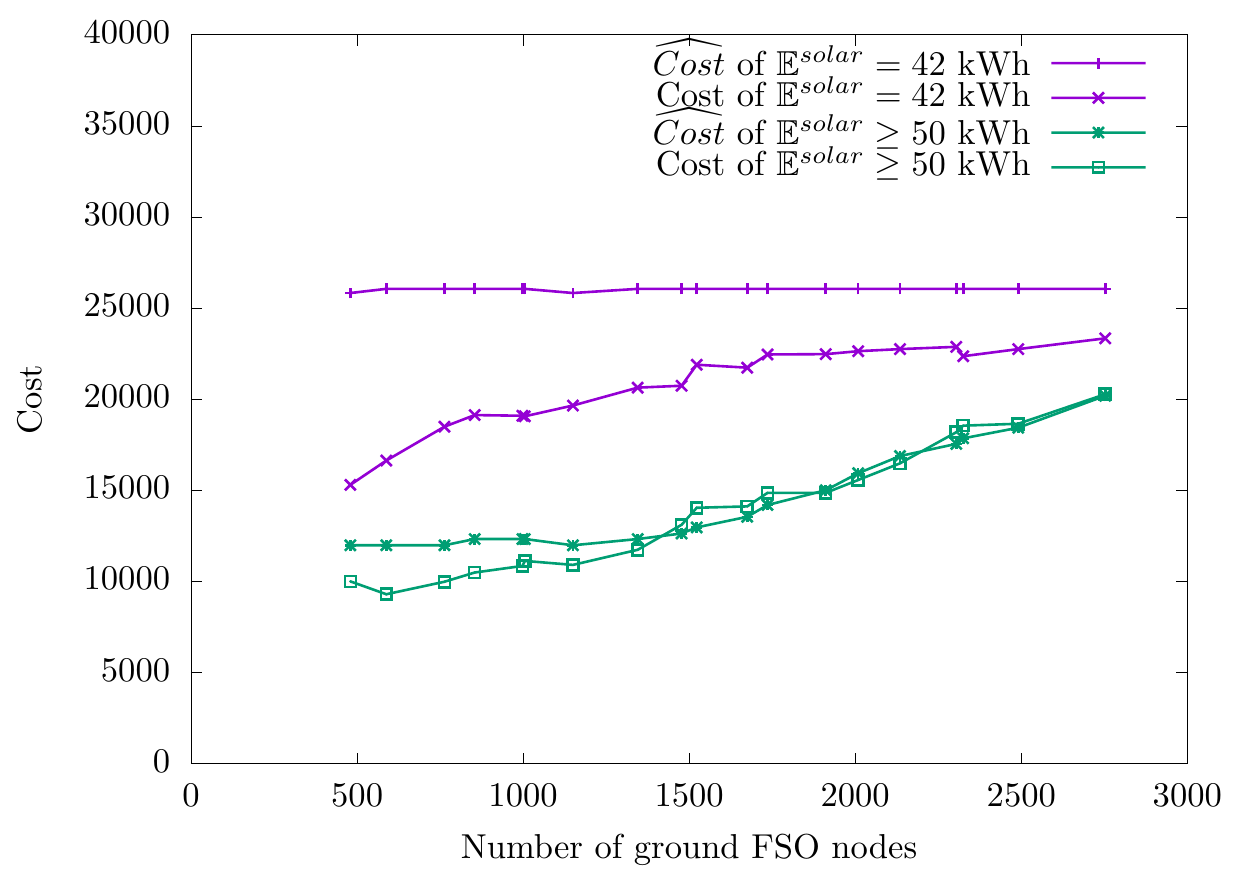}}
    \subfigure[W=80]{\includegraphics[width=0.47\textwidth]{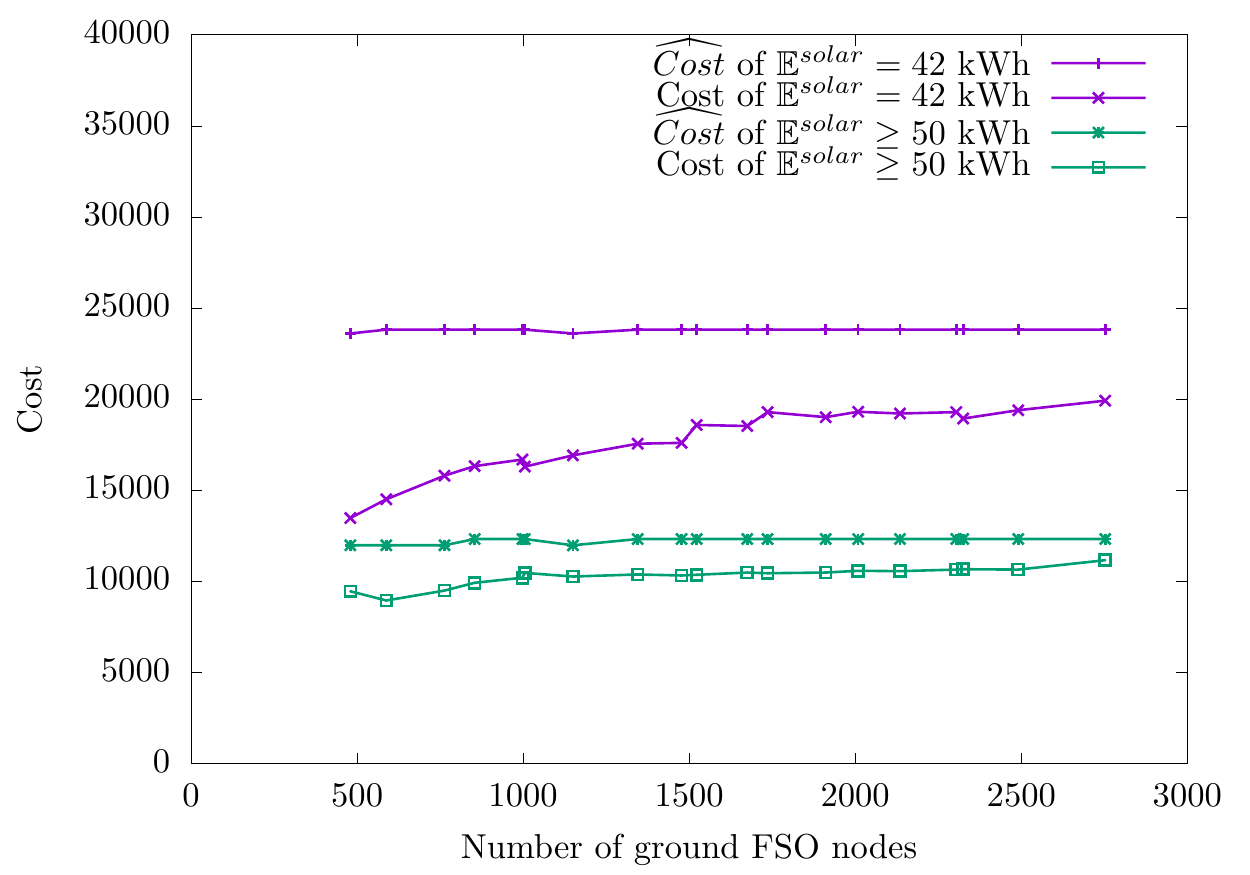}}
    \caption{Real costs and overestimated costs with $\W=40$ and $\W=80$.}
    \label{fig:cost}
\end{figure}

\begin{figure}[tbh]
    \centering
    \subfigure[W=40]{\includegraphics[width=0.48\textwidth]{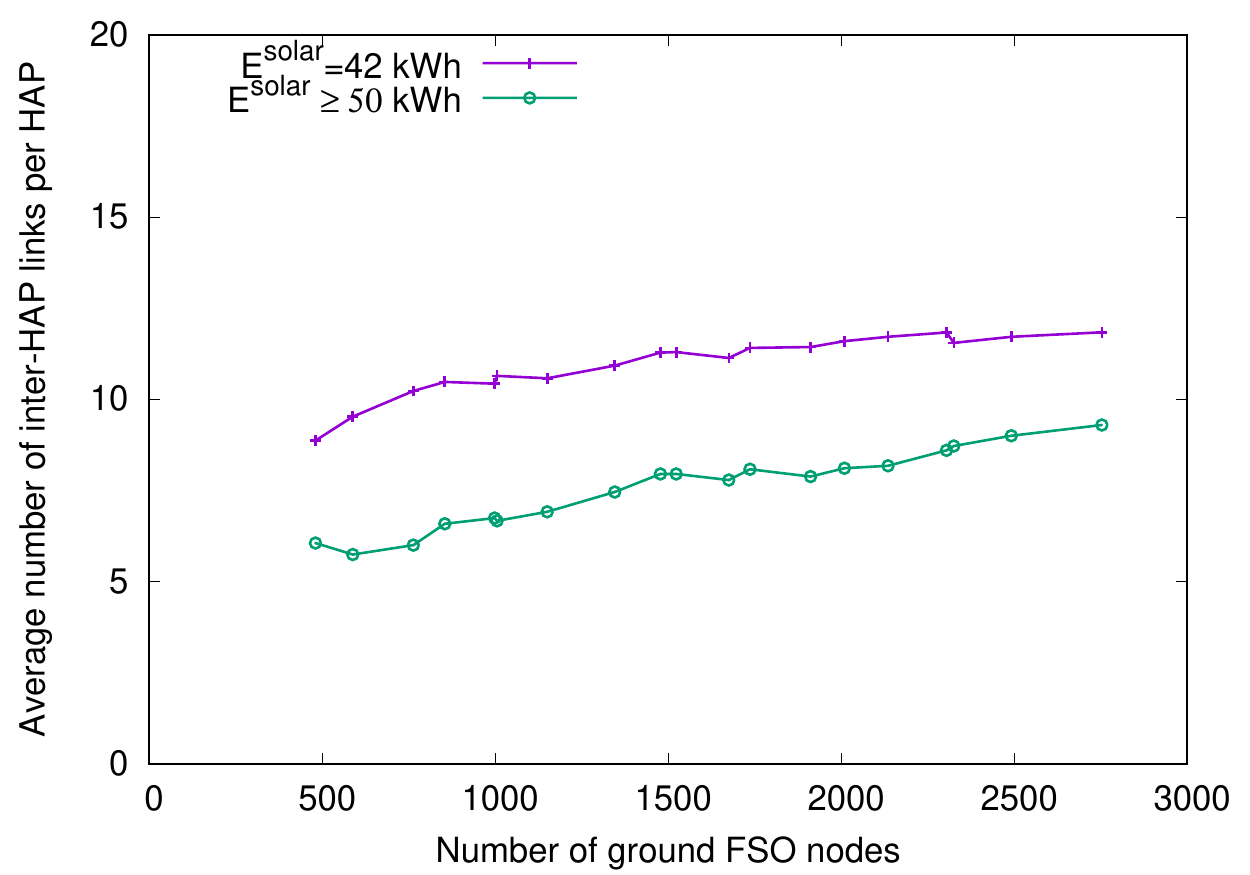}}
    \subfigure[W=80]{\includegraphics[width=0.48\textwidth]{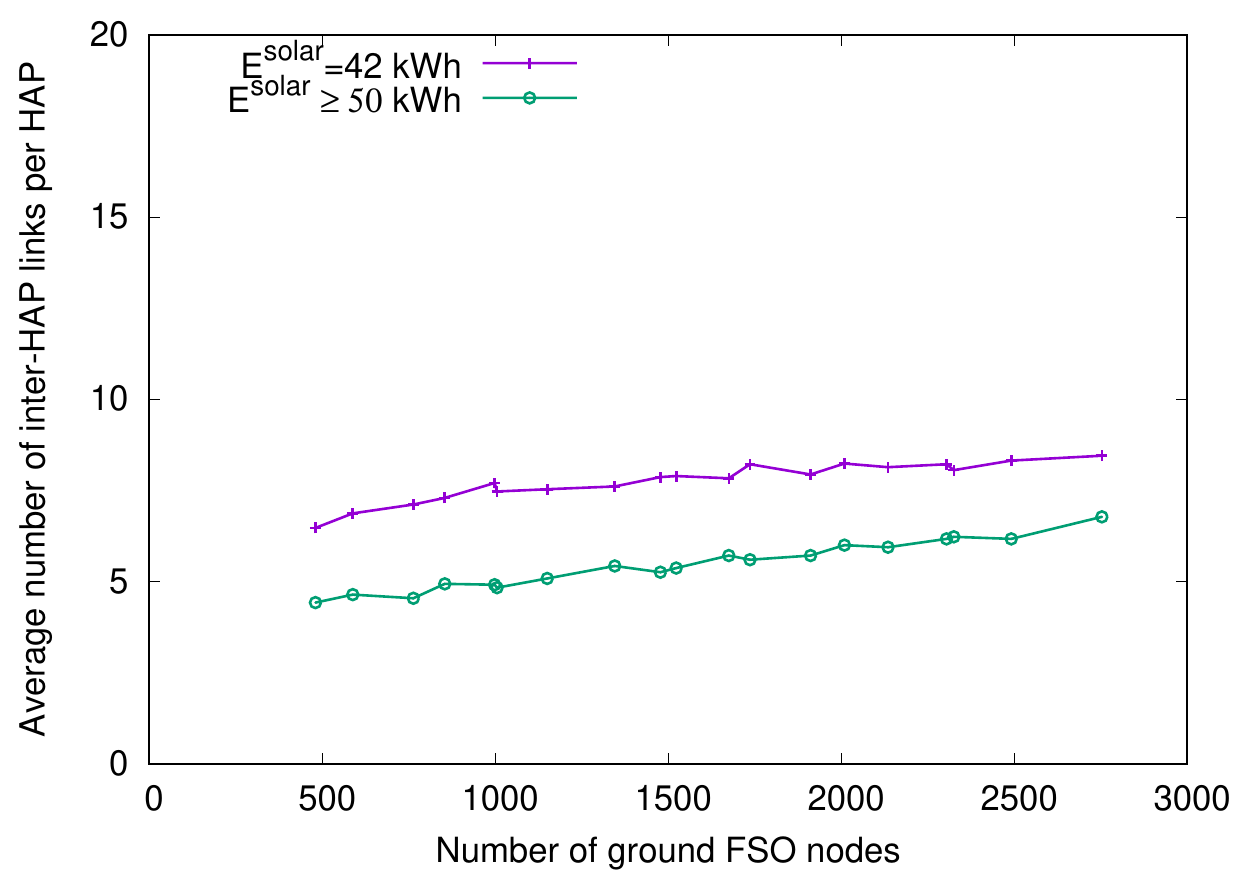}}
    \caption{Number of inter-HAP links per HAP when (a) $\W=40$ and (b) $\W=80$ for different solar energy levels.}
    \label{fig:degree-HAP}
\end{figure}

\subsection{Quality of cost estimation}

Figure \ref{fig:cost} presents the estimated and actual costs for different solar energy levels and wavelength densities. The estimated cost was very close to the actual cost, mostly for $\Psolar \geq 50$kWh and $\W=40$. 

Parameter $\V$, the threshold of the number of inter-HAP links of a HAP, affects the quality of the cost estimation. To evaluate the choice of $\V$, we compared it with the number of inter-HAP links that a HAP finally has. Figure \ref{fig:degree-HAP} shows the average number of inter-HAP links per HAP. When there were 40 wavelengths per link, the average number of inter-HAP links per HAP varied between 5.7 and 9.3 for $\Psolar \geq 50$ kWh and $\V=10$, and between 8.8 and 11.8 for $\Psolar=42$ kWh while $\V$ raised up to 12. Hence, the value of $\V$ was close to the actual number of inter-HAP links required by a HAP. However, when there were 80 wavelengths per link, the average number of Inter-HAP links per HAP was reduced to between 4.4 and 8.4, which is slightly far from the threshold $\V=10$. A smaller $\V$ may help better estimate of the optimal cost in these cases.

\section{Conclusions}
\label{sec:conclusions}

Using \msupp~configuration widens a HAP footprint, however, its application is constrained by the available solar energy of the HAP. Moreover, \msupp~configuration may imply an extra investment cost due to additional \sFSO~transceivers in comparison with single FSO transceiver configuration. This study focused on determining the optimal \msupp~configuration. First, we proposed a set of closed-form expressions for computing the coverage of an \msuppconf~in terms of beam widths of the principal and supplementary transceivers and number of supplementary FSO transceivers. Second, we proposed an algorithm to determine the optimal \msupp~configuration that minimizes the total HAP network cost. Third, we designed a HAP network topology using the optimal configuration to achieve a minimal final cost.

The simulation results showed that \msupp~significantly extended the HAP footprint. With the testing dataset, the extended footprint radii were generally two times larger than the single FSO transceiver footprint radii, leading to a four-fold larger coverage surface. The network cost with the optimal \msupp~configuration was as low as 54\% of the network cost when using a single \sFSO~transceiver on a HAP.

\section*{Acknowledgements}
This research was funded by the Vietnam National Foundation for Science and Technology Development (NAFOSTED) under grant number 102.02-2018.305.

\bibliographystyle{ieeetr}
\bibliography{FSO-HAP}

\appendix 
\section{Proof of Lemma 1}
\label{proof-lemma1}
\begin{proof}
Let $x=\cos(\alpha/2)$, $a=\sigma \Hh$, and $b= \frac{P_{tx}\Rrx^2}{2 \Hh^2}$ then 
\begin{equation}
P^{rx}_j (x) =  e^{-a/ x} \frac{ b x^2}{ (1-x)}
\end{equation}
Calculate the derivative of $P^{rx}_j (x)$ we get
\begin{equation}
P^{'rx}_j (x) =  e^{-a/ x} \left( \frac{a}{1-x} + \frac{2x-x^2}{(1-x)^2}\right)b
\end{equation}
Thus, the derivative of $P^{rx}_j (\alpha)$ is
\begin{equation}
P^{'rx}_j (\alpha) =  P^{'rx}_j (x). (-\sin(\alpha))
\end{equation}
Beam $\alpha$ is limited between  $[0 .. \pi]$ because it orients to the ground. Thus, $x \in [0 .. 1]$. Consequently, $1-x>0$ and $2x-x^2>0$. In addition, $a,b>0$, then $P^{'rx}_j (x)>0$ for all $x \in [0 .. 1]$. 
Because $-\sin(\alpha)<0, \forall \alpha \in [0 .. \pi]$, thus, $P^{'rx}_j(\alpha) <0$. Consequently, $P^{rx}_j(\alpha)$ decreases with $\alpha$. 
\end{proof}
\section{Calculation of extended coverage radius of \msupp~configuration} 
\label{sec:calculation}
This section identifies formulas that calculate the extended coverage radius of an \msupp~configuration characterized by the principal beam width $\alpha$, supplementary beam width $\beta$ and number of supplementary beams $m$. 

Conventionally, the coverage provided by a bundle of transmitters is calculated as if the transmitters project perpendicular to the ground. In \msupp~configuration, the principal beam in the center is large, and it pushes the supplementary \sFSO~transceiver projection directions far from perpendicular to the ground. These supplementary beams form oblique cones that intersect with the ground plane in ellipses. Considering of the elliptical form adds more complexity to the calculation.

\begin{figure}[tbh]
   \includegraphics[width=0.5 \textwidth]{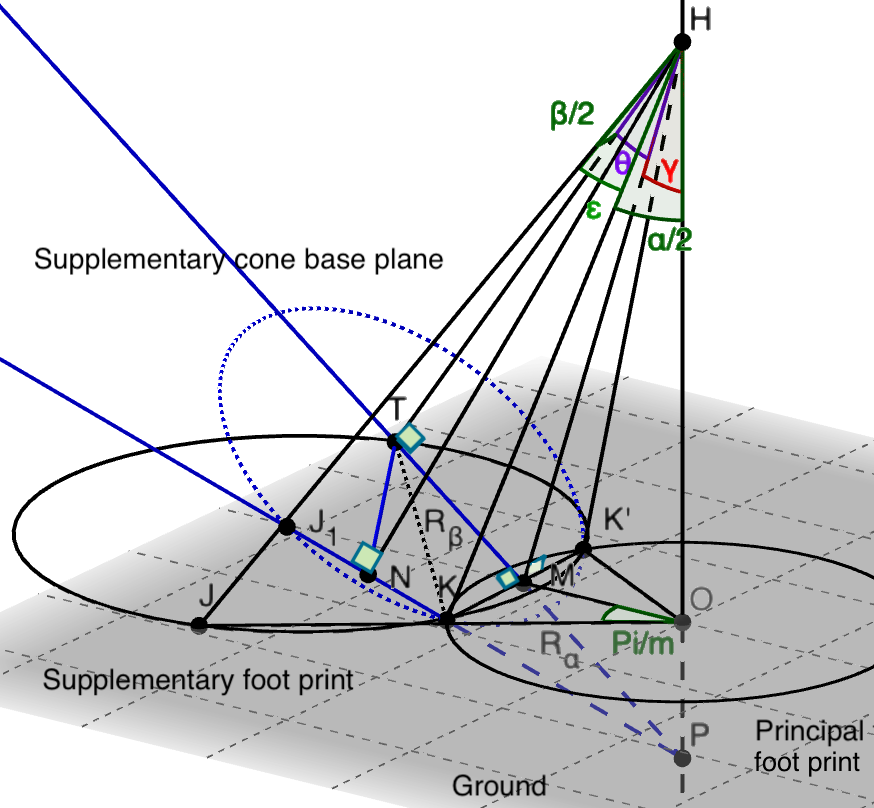}
    \caption{Computation of the distance from supplementary FSO transceivers and the border of extended coverage area $L_J$ in function of Beta.}
    \label{fig:LjversusBeta}
\end{figure}

In Figure \ref{fig:LjversusBeta}, $H$ denotes the position of a HAP, and its projection on the ground plane is $O$, thus $HO=\Hh$. The principal beam forms a right circular cone whose axis is $HO$. The cone intersects the ground plane by a circle of radius $\Ralpha$, which defines the principal footprint. The beam of a supplementary FSO transceiver is an oblique cone intersecting the ground plane by an ellipse that defines the corresponding supplementary footprint. The cone of the supplementary beam intersects with the cone of the principal beam by two lines: $HK$ and $HK'$ where $K$ and $K'$ are the two intersection points of the principal and supplementary footprints. Thus, $OK=OK'=R_{\alpha}$. 

$m$ supplementary FSO transceivers are arranged evenly around the principal transceiver, each of which is responsible for extending the coverage within an angle of $2 \pi/m$ from the center $O$. The responsible angle of the supplementary FSO transceiver in Figure  \ref{fig:LjversusBeta} is defined by rays $\overrightarrow{OK}$ and $\overrightarrow{OK'}$. Thus, $\widehat{KOK'}=2\pi/m$.

Ray $\overrightarrow{OK}$ intersects with the supplementary beam cone at $J$, then $OJ$ is the radius of the extended coverage region. Readers refer to Figure \ref{fig:mFSOconf} for a complete view of the extended coverage circle and the positions of $K$, $K'$ and $J$ on the ground.

Since the principal beam width is $\alpha$, then $\widehat{OHK}=\alpha/2$.

Let the base plane containing $K$ and $K'$ of the supplementary beam cone cuts the cone axis at $T$, the primary cone axis $HO$ at $P$, and $HJ$ at $J_1$. Then $\widehat{THK}=\beta/2$. In addition, the supplementary cone intersects with this base plane by a circle containing $K, K'$ with center $T$. Let $R_{\beta}$ be the radius of the circle, then $TK=TK'=R_{\beta}$.

Let $M$ be the midpoint of $KK'$ then $H,O,T,M$ belong to the same plane. 

Let $\xi=\widehat{KHJ}$. The extended coverage radius is $\Rext=OJ=HO \tan(\widehat{OHJ})=\Hh \tan(\at + \xi)$. Thus,
$$\Rext=\Hh. \tan\left(2(\axt) -\at \right)$$
\begin{equation}
\boxed{
\Rext=\Hh \frac{2\tan(\axt) -\tan(\at) (1- \tan^2(\axt))}{1 - \tan^2(\axt) + 2 \tan(\axt).\tan(\at)} 
}
\label{eq:r-extended}
\end{equation}

\subsection{Calculation of $\tan(\axt)$ }

Let $N$ be the midpoint of $KJ_1$. As $K$ and $J_1$ are at the intersection of the supplementary cone and its base plane, $HK=HJ_1$, $HN \perp KJ_1$, and $HN$ is the angle bisector of $\widehat{KHJ_1}$. Therefore, $\widehat{NHK} =\xi/2$, thus $\widehat{NHP} =\axt$. In addition, since $KO$ is on the base plane of the principal cone, $HO \perp KO$. Thus, $\triangle PNH$ and $\triangle POK$ are similar right triangles. Consequently, $ \widehat{OKP}=\widehat{NHP}=\axt$. Furthermore,
\begin{equation}
\tan(\frac{\xi + \alpha}{2})= \frac{OP}{OK}= \frac{OP}{R_{\alpha}}
\label{eq:tan-xi-alpha}
\end{equation}
Let $\widehat{OHM}= \gamma$ and $\widehat{THM}= \theta$
Then $\widehat{OHT}=\theta+\gamma$. 

Because $MO$ is on the base plan of the principal cone, $MO \perp HO$.  In addition, as  $PT$ is on the base plane of the supplementary cone whose axis is $HT$ then $HT \perp PT$. Consequently, $\triangle PTH$ and $\triangle POM$ are similar right triangles. We can deduce that $\widehat{PMO}=\widehat{PHT}=\theta+\gamma$. Therefore,
\begin{equation}
\tan(\theta +\gamma) =\frac{OP}{OM}=\frac{OP}{R_{\alpha}.\cos(\pim)} \nonumber
\label{eq:tan-theta-gamma}
\end{equation}
Combining with \eqref{eq:tan-xi-alpha} we deduce :
\begin{equation}
\tan(\frac{\xi + \alpha}{2})=\tan(\theta + \gamma ). \cos(\pim) 
\label{eq:tan-xi-alpha1}
\end{equation}
Thus
\begin{equation}
\boxed{
   \tan(\frac{\xi + \alpha}{2}) =\frac{\tan(\gamma) + \tan(\theta)}{1-\tan(\gamma).\tan(\theta)}.\cos(\pim)
  }
\label{eq:tan-xi-alpha2}
\end{equation}
Since $ \gamma=\widehat{OHM}$ then, $\tan(\gamma)=\frac{MO}{HO}$. \\
From right triangle $\triangle OMK$ we have $MO= OK.\cos(\pim)$.\\
From right triangle $\triangle HOK$ we have $HO= OK/\tan(\at)$. \\
Thus
\begin{equation}
\boxed{
\tan(\gamma)=\tan(\at).\cos(\pim)
}
\label{eq:gamma}
\end{equation}
It remains to calculate $\tan{(\theta)}$.

\subsection{Calculation of $\tan{(\theta)}$} 
Look at the right triangle $\triangle HTM$, we can see that:
\begin{equation}
\tan(\theta)=\frac{TM}{TH} 
\label{eq:tan-theta}
\end{equation}

Since $K$ and $K'$ are on a circle centered at $T$, and $M$ is the midpoint of $KK'$ then  $\triangle TMK$ is a right triangle, then 
\begin{equation}
TM =\sqrt{TK^2-KM^2}=\sqrt{R_{\beta}^2 - R_{\alpha}^2. \sin^2(\pim)}
\label{eq:TM}
\end{equation}
Easy to find that $\triangle THK$ is another right triangle then 
\begin{equation}
TH= TK / \tan(\bt) = R_{\beta}/ \tan(\bt)
\label{eq:TH}
\end{equation}
Replacing \eqref{eq:TM} and \eqref{eq:TH} in to \eqref{eq:tan-theta} we get
\begin{eqnarray}
\tan(\theta)&=&\frac{\sqrt{R_{\beta}^2 - R_{\alpha}^2.\sin^2(\pim)}}{R_{\beta}/ \tan(\bt)} \nonumber \\
		&=& \tan(\bt) \sqrt{1 - (\frac{R_{\alpha}}{R_{\beta}})^2.\sin^2(\pim)} 
		\label{eq:tan-theta2}
\end{eqnarray}
From right triangle $\triangle HTK$ we obtain $R_{\beta}=HK\sin(\bt)$. \\
From right triangle $\triangle HOK$ we obtain $R_{\alpha}= HK \sin(\at)$.\\
Replacing these values to \eqref{eq:tan-theta2}, we obtain:
\begin{equation}
\boxed{
\tan(\theta)=\frac{\sqrt{\sin^2(\bt)  -\sin^2(\at).\sin^2(\pim)}}{\cos(\bt)} 
\label{eq:tan:theta3}
}
\end{equation}

Substituting the values of $\tan({\gamma})$ in \eqref{eq:gamma} and $\tan({\theta})$ in \eqref{eq:tan:theta3} into \eqref{eq:tan-xi-alpha2}, we obtain $\tan({\axt})$. Subsequently, replacing the obtained $\tan(\axt)$ to \eqref{eq:r-extended} we get $\Rext$.

\end{document}